\documentclass[11pt]{article}
\usepackage{amsthm,amsmath,amsfonts,amssymb,mathbbol}
\pdfoutput=1
\usepackage{bm}
\usepackage{textcomp}
\usepackage{graphics,color}
\usepackage[colorlinks,citecolor=blue,urlcolor=blue]{hyperref}
\usepackage{dsfont}
\usepackage{graphicx}
\usepackage{algorithm}
\usepackage{algpseudocode}
\usepackage{adjustbox}
\usepackage{subcaption} 
\usepackage{float}
\usepackage{placeins}

\usepackage{times,mathptm}

\usepackage[round]{natbib}
\usepackage{booktabs}
\usepackage{multirow}

\newcommand{\bx}{{\bm{x}}}
\newcommand{\bX}{{\bm{X}}}
\newcommand{\bZ}{{\bm{Z}}}

\newcommand{\bI}{{\bm{I}}}
\newcommand{\bn}{{\bm{n}}}

\newcommand{\bbeta}{{\bm{\beta}}}

\newcommand{\E}{{\mathrm{E}}}
\newcommand{\pr}{{\mathrm{P}}}

\newcommand{\mbL}{{\mathbb{L}}}
\newcommand{\mbX}{{\mathbb{X}}}

\newcommand{\mbS}{{\mathbb{S}}}

\newcommand{\var}{\operatorname{Var}}

\newcommand{\equald}{\overset{d}{=}}
\newcommand{\ind}{\perp\!\!\!\!\perp} 

\newtheorem{theorem}{Theorem}[section]
\newtheorem{proposition}[theorem]{Proposition}
\newtheorem{corollary}[theorem]{Corollary}
\newtheorem{remark}[theorem]{Remark}

\title{Estimation of Treatment Harm Rate via Partitioning}
\author{Wei Liang and Changbao Wu}
\date{} 

\begin{document}

\maketitle 
\renewcommand{\thefootnote}{}
\footnotetext{Emails. Wei Liang: \texttt{w62liang@uwaterloo.ca}, Changbao Wu: \texttt{cbwu@uwaterloo.ca}}

\begin{abstract}
In causal inference with binary outcomes, there is a growing interest in estimation of treatment harm rate (THR), which is a measure of treatment risk and reveals treatment effect heterogeneity in a subpopulation. The THR is generally non-identifiable even for randomized controlled trials (RCTs), and existing works focus primarily on the estimation of the THR under either untestable identification or ambiguous model assumptions. We develop a class of partitioning-based bounds for the THR based on data from RCTs with two distinct features: Our proposed bounds effectively use available auxiliary covariates information and the bounds can be consistently estimated without relying on any untestable or ambiguous model assumptions. Finite sample performances of our proposed interval estimators along with a conservatively extended confidence interval for the THR are evaluated through Monte Carlo simulation studies. An application of the proposed methods to the ACTG 175 data is presented. A Python package named \texttt{partbte} for the partitioning-based algorithm has been developed and is available on \href{https://github.com/w62liang/partition-te}{https://github.com/w62liang/partition-te}. 
\end{abstract}


\newpage
\section{Introduction}
\label{sec:intro}
Understanding treatment effect heterogeneity (TEH) is of substantial practical importance in causal inference. Subgroup analysis and the estimation of conditional average treatment effects are two commonly used strategies for this critical issue. However, these approaches primarily focus on difference-in-means metrics, which may fail to fully capture the extent of TEH by overlooking heterogeneity that exists in the subpopulation \citep{zhang2013assessing,kallus2022s}. For instance, while a medication may yield a positive average effect for a particular subpopulation, it could still pose risks to certain individuals within that group --- some may experience adverse effects despite the overall benefit. To obtain a more comprehensive understanding of TEH, it is necessary to consider the joint distribution of counterfactual outcomes, which is encapsulated by the treatment harm rate (THR) in a randomized controlled trial (RCT) when the counterfactual outcomes are binary. The THR is defined as the proportion of individuals negatively affected by a treatment in a given population and reflects the average risk in a population associated with the treatment. Exploring the THR complements existing research on TEH and is increasingly important in real-world applications.

The THR is generally nonidentifiable even in a RCT, unless we impose some strong conditions on the data structure such as conditional independence \citep{zhang2013assessing,shen2013treatment} or joint distribution restrictions \citep{huang2019constructing}. \citet{gadbury2004individual} studied the partial identifiability and derived the sharp bounds (Fr\'echet-Hoeffding bounds) of the THR in RCTs without covariates. The Fr\'echet-Hoeffding bounds of the THR were later extended by \citet{zhang2013assessing} and \citet{kallus2022s} to observational studies in the presence of covariates. \citet{wu2024quantifying} further improved the bounds assuming positive conditional correlation between the two potential outcomes and developed a sensitivity analysis framework for the assumption. 

In RCTs, it is common to report the estimated average treatment effect of the clinical subjects, which has been justified at the design stage by the randomization on treatment assignments. However, there are two important considerations on the estimators of the THR that should also be reported for RCTs. \textbf{First, auxiliary information should be effectively incorporated in the estimation of the THR to reduce the conservativeness of the estimators.} For example, the interval estimator provided by \citet{gadbury2004individual} might be too wide to be practically informative. The width of the interval estimator can be substantially reduced  by effective use of the auxiliary covariates information. \textbf{Second, justification of the estimators of the THR after covariate adjustment should not involve any untestable identification or ambiguous model assumptions other than treatment randomization}. For instances, \citet{huang2019constructing} required the users to impose zero proportion restrictions on some potential outcome pairs in order to identify the THR; \citet{zhang2013assessing} and \citet{shen2013treatment} assumed that the two potential outcomes are conditionally independent given the covariates. These identification assumptions on the joint distribution of the potential outcomes are un-testable and conclusions based on which are thus unreliable. One might argue that sensitivity analyses can be incorporated to address issues of assumption violations, but choosing a reasonable range of the sensitivity parameter brings in another challenging issue in the absence of sufficient background knowledge. 

In this paper, we develop a class of partitioning-based bounds for the THR by taking into account of the two aforementioned considerations. Our method is motivated from a key observation that the sharp bounds of the THR can be attained under a partition of the covariates space with at most four cells and the quantity of interest in each partition can be consistently estimated with RCT data. We show that the optimal partitioning rule is the Bayes classifier for a weighted multiclass classification problem. Probabilistic classification algorithms are employed to estimate the nuisance parameters and to form the partitioning. The proposed interval estimators can be highly informative if the underlying models capture the true data patterns, which are in a spirit similar to the model-assisted estimators in survey sampling and causal inference \citep{wu2001model,wu2003optimal,wu2020sampling,ye2023toward}. 

The rest of this paper is organized as follows. In section \ref{sec: notation}, we derive the partitioning-based bounds for the THR and show that a partition of four cells can yield the sharp bounds of the THR under standard assumptions used in causal inference literature. In section \ref{sec:nonpara-estimation}, we show how to estimate the partitioning-based bounds given a partition with RCT data. We discuss in section \ref{sec:space-partition} on how to create a partition by plugging in probabilistic classifiers and how to create pseudo external data by cross-fitting. Confidence intervals for the partitioning-based bounds, which can be computed numerically based on the asymptotic distributions, are provided in section \ref{sec:precision-evaluation} for the evaluation of the estimated bounds. In section \ref{sec:numerical}, we report results from Monte Carlo simulation studies to assess the finite-sample performance of the proposed interval estimators. An application of the proposed estimators to analyzing the AIDS Clinical Trials Group Study 175 is presented in section \ref{sec:real-data}, and some concluding remarks are given in section \ref{CR}.  Proofs and technical details as well as additional simulation results are relegated to the Appendix. 

\section{Partitioning-Based Bounds}
\label{sec: notation}

Suppose we are interested in the effect of a treatment of two levels, $a\in\{0,1\}$, on individuals from an infinite super population. We denote by $Y(a)$ the potential outcome given the treatment $a$ according to the Neyman-Rubin potential outcome framework \citep{rubin1974estimating} and assume $Y(a)$ takes values in $\{-1,1\}$, with $Y(a)=1$ being favorable. Let $A\in\{0,1\}$ be the treatment indicator, $Y=Y(A)=AY(1)+(1-A)Y(0)$ the observed outcome, $\bX$ a vector of pre-treatment baseline covariates, and $\mathbb{X}$ the support of $\bX$. We consider the ``unconfoundedness assumption'' that $Y(1),Y(0)\ind A\mid\bX$ and the ``overlap assumption'' that $0<\pr(A=1\mid\bX)<1$ over $\mbX$. The unconfoundedness assumption is un-testable, but it holds naturally in randomized controlled trials (RCTs). We consider randomized designs in following sections for estimation and thus the above two identification assumptions are only the focus of the current section on more general results. The parameter of interest is the treatment harm rate (THR), defined as $\theta=\pr(Y(0)=1,Y(1)=-1)$. 

The parameter $\theta$ is generally non-identifiable even in RCTs unless $\var(Y\mid A=1,\bX)=0$ or $\var(Y\mid A=0,\bX)=0$ almost surely \citep{kallus2022s}. However, the $\theta$ can be partially identified under the above assumptions with the sharp bounds given by $\theta\in[L(\theta),U(\theta)]$ where $L(\theta)=\E[\max\{0,\mu_{0}(\bX)-\mu_{1}(\bX)\}]$, $U(\theta)=\E[\min\{\mu_{0}(\bX),1-\mu_{1}(\bX)\}]$, $\mu_{a}(\bX)=\pr(Y(a)=1\mid\bX)$ for $a=0,1$ \citep{zhang2013assessing,kallus2022s,wu2024quantifying}.

We consider a class of partitioning-based bounds (PBBs) for $\theta$. Given an event $\mbX_j$, we define $\mu_a(\mbX_j)=\pr(Y(a)=1\mid\bX\in \mbX_j)$ for $a=0,1$. Consider a partition of the support of $\bX$ as $\Delta=\{\mbX_1,\cdots,\mbX_J\}$ so that $\mbX=\bigcup_{j=1}^J\mbX_j$ and $\mbX_j$ are disjoint of each other (We assume $\mbX_j$ are all measurable). Based on the partition $\Delta$, we can decompose $\theta$ into $\theta=\sum_{j=1}^J\rho_j\theta_j$ where $\theta_j=\pr(Y(0)=1,Y(1)=-1\mid\bX\in{\mbX}_j)$ and $\rho_j=\pr(\bX\in\mbX_j)$. Each $\theta_j$ is bounded by  $\max\{0,\mu_0(\mbX_j)-\mu_1(\mbX_j)\}\leq\theta_j\leq \min\{\mu_0(\mbX_j),1-\mu_1(\mbX_j)\}$ due to the Fr\'echet-Hoeffding inequality \citep{frechet1935generalisation}. Then, the PBBs pair of $\theta$ under the partition $\Delta$ is given by $B_{\theta}(\Delta)=[L(\theta,\Delta), U(\theta,\Delta)]$ where $L(\theta,\Delta) = \sum_{j=1}^J\rho_j \max\{0,\mu_0(\mbX_j)-\mu_1(\mbX_j)\}$ and $U(\theta,\Delta) = \sum_{j=1}^J\rho_j \min\{\mu_0(\mbX_j),1-\mu_1(\mbX_j)\}$. 

Two trivial partitions are $\Delta_{0}=\{\mbX\}$ and $\Delta_{\infty}=\bigcup_{\bx\in\mbX}\{\{\bx\}\}$, the latter of which is a collection of all singleton sets. Let us define $\sum_{x\in\mbX}\pr(\bX=\bx)h(\bx)=\int_{\mbX}h(\bx)f_{\bX}(\bx)\nu(d\bx)$ for any function $h:\mbX\rightarrow \mathbb{R}$ where $f_{\bX}(\bx)$ is the density of $\bX$ with respect to the Lebesgue measure $\nu$. Then, $B_{\theta}(\Delta_{0})$ is the sharp bounds under the assumption $Y(1),Y(0)\ind T$ without covariates, first derived by \citet{gadbury2004individual}, and $B_{\theta}(\Delta_{\infty})=[L(\theta),U(\theta)]$ refers to the sharp bounds of $\theta$ in the presence of covariates under the unconfoundedness and overlap assumptions. Nonetheless, it may not be surprising that, $B_{\theta}(\Delta_{\infty})$ can be attained given a partition of $\mbX$ with four cells. Furthermore, to attain either the sharp upper bound or the sharp lower bound of $\theta$, a partition with two cells is sufficient.

\begin{proposition}
{ Let $\mbL=\{0,1\}\times\{-1,1\}$ denote the value space of $(A,Y)$. The partition $\Delta_{full}=\{\mbX_{a,t},(a,t)\in\mbL\}$ yields the sharp bounds of $\theta$, i.e., $B_{\theta}(\Delta_{full})=B_{\theta}(\Delta_{\infty})$, where $\mbX_{a,t}=\{\bx\in\mbX: (1-t)/2+t\mu_{a}(\bx)\geq \mu_{max}(\bx)\}$, $\mu_{max}(\bx)= \max_{(a,t)\in\mbL}\{(1-t)/2+t\mu_{a}(\bx)\}$. Ties can be broken randomly. Because $\mbX_{a,t}$ could be an empty set, the size of $\Delta_{full}$ ranges from 1 to 4. Note that $(1-t)/2+t\mu_{a}(\bx)=\pr(Y(a)=t\mid\bX=\bx)$.
\label{prop:fullpartition}
}
\end{proposition}


\begin{proposition}
{\it Let $\mbX_{L,t}=\{\bx\in\mbX: t\mu_{0}(\bx)\leq t\mu_1(\bx)\}$ and $\mbX_{U,t}=\{\bx\in\mbX: t\mu_{0}(\bx)+t\mu_{1}(\bx)\leq t\}$ for $t\in\{-1,1\}$. Consider the partitions $\Delta_{L,full}=\{\mbX_{L,1},\mbX_{L,-1}\}$ and $\Delta_{U,full}=\{\mbX_{U,1},\mbX_{U,-1}\}$. Then, we have $L(\theta,\Delta_{L,full})=L(\theta)$ and $U(\theta,\Delta_{U,full})=U(\theta)$. 
\label{prop:fullpartition_lower_upper}
}
\end{proposition}

Let $g^*(\bx)=\arg\max_{(a,t)\in\mbL}\{(1-t)/2+t\mu_{a}(\bx)\}$ be the optimal partitioning rule which yields $\Delta_{full}$. Then, the $g^*(\bx)$ is essentially a Bayes classifier for a weighted multiclass classification problem.

\begin{proposition}
{ Let us take the partitioning as a multiclass classification problem where the label for each observation is $(A,Y)\in\{(1,1),(1,-1),(0,1),(0,-1)\}$, or equivalently, $(Y(1),Y(0))\in\{(1,?),(-1,?),(?,1),(?,-1)\}$. Consider the classifier $g:\mbX\rightarrow\mbL$. Let $m_a(g)=\pr(g(\bX)\neq (a,Y(a)))$ be the misclassification rate for individuals whose first label is equal to $a$ and $\pi_a=\pr(A=a)$. Then, $g^*(\bx)$ minimizes $m_1(g)+m_0(g)$, while the ordinary multiclass Bayes classifier, $g^B(\bx)=\arg\max_{(a,t)\in\mbL}\pr(A=a,Y=t\mid\bX=\bx)$, minimizes $\pr(g(\bX)\neq (A,Y))=\pi_1m_1(g)+\pi_0m_0(g)$.
\label{prop:partition-bayes}
}
\end{proposition}

The PBBs of $\theta$ own a very attractive property that $B_{\theta}(\Delta)$ can always be consistently estimated under the randomization of treatment assignments as long as the size of the partition does not grow too large, as will be discussed in the next section. Whether the bounds are informative relies on the provided partition. Proposition \ref{prop:fullpartition} and \ref{prop:partition-bayes} provide a guideline on how to find the optimal partition that yields the sharp bounds. The second key observation in this paper is that finer partitioning cannot hurt the sharpness of the bounds.

\begin{proposition}
{\it A partition $\Delta_1$ is said to be finer than a partition $\Delta_2$ if and only if for every cell $\omega_2\in\Delta_2$, there exists a cell $\omega_1\in\Delta_1$ such that $\omega_2\subseteq\omega_1$. Then for two partitions $\Delta_1$ and $\Delta_2$, we have $B_{\theta}(\Delta_1)\subseteq B_{\theta}(\Delta_2)$ if $\Delta_1$ is finer than $\Delta_2$.
\label{prop:finerpartition}
}
\end{proposition}


\begin{remark}
    Due to Proposition \ref{prop:fullpartition_lower_upper} and \ref{prop:finerpartition}, the partition, denoted by $\Delta_{full,L}\times \Delta_{full,U}=\{\mbX_{L,s}\cap\mbX_{U,t},s,t\in\{1,-1\}\}$, as pairwise intersections of $\Delta_{L,full}$ and $\Delta_{U,full}$ must yield sharp bounds of $\theta$. Indeed, it can be shown that $\Delta_{full,L}\times \Delta_{full,U}=\Delta_{full}$. 
\end{remark}

Some bounds are informative and some are not. For example, the bounds $[0,1]$ are totally non-informative for $\theta$. We define the information (for $\theta$) carried by $\Delta$, or equivalently the information of $B_{\theta}(\Delta)$, as one minus the width of $B_{\theta}(\Delta)$, i.e.,
\[
I_{\theta}(\Delta) = 1+L(\theta,\Delta) -U(\theta,\Delta)=\sum_{j=1}^J\rho_j\max\{\mu_{1}(\mbX_j),1-\mu_{1}(\mbX_j),\mu_{0}(\mbX_j),1-\mu_{0}(\mbX_j)\}.
\]
The sharper the $B_{\theta}(\Delta)$ is, the more information the partition $\Delta$ carries. Because $B_{\theta}(\Delta_{\infty})$ is sharp in the presence of covariates, the information upper bound for $\theta$ will be $I_{\theta}(\Delta_{\infty})=I_{\theta}(\Delta_{full})=\E[\max\{\mu_{0}(\bX),1-\mu_{0}(\bX),\mu_{1}(\bX),1-\mu_{1}(\bX)\}]$.  Proposition \ref{prop:finerpartition} implies that we will not lose information by finer partitioning, which is an important property for the technique we will discuss in section \ref{subsec:cross-fit}.


\section{Non-Parametric Estimation}
\label{sec:nonpara-estimation}

Consider a sample of individuals indexed in $\mbS=\{1,\cdots,n\}$. Let $\{(Y_i(1),Y_i(0),\bX_i),i\in\mbS\}$ be the attributes of the sample units as $n$ i.i.d. copies of $(Y(1),Y(0),\bX)$. Let $A_i$ be the treatment assigned to the $i$-th unit. We assume that the observed data are RCT data, identified by the following two assumptions: 
\begin{enumerate}
    \item[A1.] $\{A_1,\cdots,A_n\}\ind \{(Y_i(1),Y_i(0),\bX_i)\}_{i=1}^n$.
    \item[A2.] $\E(A_i)=\pi_1\in(0,1)$ for $i=1,\cdots,n$.
\end{enumerate}
Let $Y_i=Y_i(A_i)$ denote the observed outcome and $\mathcal{S}=\{(Y_i,A_i,\bX_i),i\in\mbS\}$ the observed sample data. The goal of this section is to estimate $B_{\theta}(\Delta)$ provided a partition $\Delta$ and the sample data $\mathcal{S}$ under simple randomization. We propose to estimate $\mu_a(\mbX_j)$ nonparametrically by
\[
\hat{\mu}_{a}(\mbX_j) = \frac{\#\{i\in\mbS: Y_i=1,A_i=a,\bX_i\in \mbX_j\}}{\#\{i\in\mbS: A_i=a,\bX_i\in \mbX_j\}}.
\]
Because of simple randomization, $\hat{\mu}_{a}(\mbX_j)$ is an unbiased and consistent estimator of ${\mu}_a(\mbX_j)$. Let $n_j=\#\{i\in\mbS:\bX_i\in\mbX_j\}$ be the size of the sample units in $\mbX_j$. Then, the $L(\theta,\Delta)$ can be estimated as $\widehat{L}(\theta,\Delta)=\sum_{j=1}^Jn_j/n\max\{0,\hat{\mu}_{0}(\mbX_j)-\hat{\mu}_{1}(\mbX_j)\}$, and $U(\theta,\Delta)$ can be estimated as $\widehat{U}(\theta,\Delta)=\sum_{j=1}^Jn_j/n\min\{\hat{\mu}_{0}(\mbX_j),1-\hat{\mu}_{1}(\mbX_j)\}$. The partitioning-based interval estimator of $\theta$ (under the partition $\Delta$) is given by $\hat{B}_{\theta}(\Delta)=[\widehat{L}(\theta,\Delta),\widehat{U}(\theta,\Delta)]$. 
\begin{proposition}
\label{prop:nonpara-est}
{\it Let $\eta_n=\sum_{j=1}^J\sum_{a=0}^1\sigma_{Y(a),j}\sqrt{{\rho_j}/{{n\pi_a}}}+1/\sqrt{n}$ where $\sigma^2_{Y(a),j}=\var(Y(a)\mid\bX\in\mbX_j)$ and $\pi_a=\pr(A=a)$. Denote $\hat{I}_{\theta}(\Delta)=1+\widehat{L}(\theta,\Delta) -\widehat{U}(\theta,\Delta)$. Under simple randomization, provided a partition $\Delta$, we have $\widehat{L}(\theta,\Delta)={L}(\theta,\Delta)+O_p\left(\eta_n\right)$, $\widehat{U}(\theta,\Delta)={U}(\theta,\Delta)+O_p\left(\eta_n\right)$, and $\E[\hat{I}_{\theta}(\Delta)]\geq I_{\theta}(\Delta)$.}
\end{proposition}

Justification of the bounds estimators is solely randomization-based. It does not require any assumptions on the partition $\Delta$ except finite $J$ as long as $\Delta$ is fixed. Typically, we do not hope $J$ gets too large because the bias term $\eta_n$ converges more slowly as $J$ increases. Proposition \ref{prop:nonpara-est} also indicates that the non-parametric estimation procedure will not lead to a loss of information carried by $\Delta$.

\section{Space Partitioning}
\label{sec:space-partition}

In this section, we introduce a method for partitioning. To validate the non-parametric estimators in section \ref{sec:nonpara-estimation}, the partition should be fixed regarding the sample data $\mathcal{S}$. Therefore, we assume there is an external dataset $\mathcal{G}=\{(Y_i,\bX_i,A_i),i\in\mathbb{G} \}$, independent of $\mathcal{S}$, so that any partitions created using $\mathcal{G}$ can be viewed as fixed conditional on $\mathcal{G}$. The data $\mathcal{G}$ can be from observational studies where treatment assignments are non-randomized. Later in this section, we will show how to create pseudo external datasets by cross-fitting using the experimental data at hand when the required data are not available from other sources. 

\subsection{Plug-in Partition}

Let $N=|\mathbb{G}|$ be the size of be external data $\mathcal{G}$ and $\hat{\mu}_{a,N}(\bx)$ be an estimator of $\mu_a(\bx)$ trained on $\mathcal{G}$ for $a=0,1$. The $\hat{\mu}_{a,N}(\bx)$ should be independent of the sample $\mathcal{S}$. Motivated by Proposition \ref{prop:fullpartition}, we create a plug-in partition as $\Delta_{plug}=\Delta_{plug}(\hat{\mu}_{0,N}(\bx),\hat{\mu}_{1,N}(\bx))=\{\hat{\mbX}_{a,t},(a,t)\in\mbL\}$, where $\hat{\mbX}_{a,t}=\{\bx\in\mbX: (1-t)/2+t\hat{\mu}_{a,N}(\bx)\geq \hat{\mu}_{max,N}(\bx)\}$ as a guess of $\mbX_{a,t}$ by plugging in the estimators $\hat{\mu}_{1,N}(\bx)$, $\hat{\mu}_{0,N}(\bx)$, and $\hat{\mu}_{max,N}(\bx)= \max_{(a,t)\in\mbL}\{(1-t)/2+t\hat{\mu}_{a,N}(\bx)\}$. Ties will be broken randomly. Here, the functions $\mu_a(\bx)$ are referred to as nuisance parameters. They are not the parameters of primary interest. However, the information carried by $\Delta_{plug}$ does rely on whether we can accurately estimate these nuisance parameters. 

\begin{proposition} The information carried by $\Delta_{plug}$ satisfies 
\label{prop: plug-in partition}
    \[
    I_{\theta}(\Delta_{plug})\geq I_{\theta}(\Delta_{\infty}) - \E\left[\mu_{max}(\bX) - \hat{\mu}_{max,N}(\bX) + \max_{a=0,1}|\hat{\mu}_{a,N}(\bX) - \mu_a(\bX)|\mid \mathcal{G}\right],
    \]
    where $\E[\cdot\mid\mathcal{G}]$ denotes taking expectation conditional on the external data $\mathcal{G}$.
\end{proposition}

\begin{corollary}
    If $\E[|\hat{\mu}_{a,N}(\bX) - \mu_a(\bX)|\mid \mathcal{G}]=o_p(1)$ for $a=0,1$, then there must be $I_{\theta}(\Delta_{plug})\geq I_{\theta}(\Delta_{\infty}) - o_p(1)$. The proof can be easily obtained by noting that $|\mu_{max}(\bX) - \hat{\mu}_{max,N}(\bX)|\leq |\hat{\mu}_{1,N}(\bX) - \mu_1(\bX)|+|\hat{\mu}_{0,N}(\bX) - \mu_0(\bX)|$.
\end{corollary}
The information upper bound is asymptotically reached if the estimator $\hat{\mu}_{a,N}$ is consistent in the sense that $\E[|\hat{\mu}_{a,N}(\bX) - \mu_a(\bX)|\mid \mathcal{G}]=o_p(1)$ for $a=0,1$. We consider popular machine learning classification algorithms for the estimation of the nuisance parameters in the next section. 

\begin{remark}
    An alternative way to partitioning is to solve the weighted multiclass classification problem as formulated in Proposition \ref{prop:partition-bayes}, which can be realized by minimizing a surrogate loss function of the weighted misclassification risk. The topic is for future work and will not be discussed in the current paper.
\end{remark}

\subsection{Probabilistic Classifiers}
In machine learning, a classifier is a function that takes the covariates as input and the predicted label of the categorical outcome as output. A probabilistic classifier is a classifier that also outputs the estimated probability of the outcome belonging to a certain class given the covariates. Many machine learning models are probabilistic, and their predicted labels of the outcomes are essentially produced by applying a threshold to the estimated probabilities. Some models are margin-based and not inherently probabilistic such as support vector machines. With certain techniques such as Platt scaling, however, the margins to the decision boundaries can be transformed into probabilistic estimates \citep{platt1999probabilistic}. In the Appendix, we give a brief review of six classification algorithms on binary outcomes -- logistic regression, Gaussian naive Bayes, K-nearest-neighbors, support vector machines, random forests, and gradient boosting -- and how they do probabilistic estimation. These algorithms will be employed in our numerical experiments for the estimation of nuisance parameters. For more details of these algorithms, we refer readers to \citet{hastie2009elements}. While highly flexible and capable of being fast implemented in software such as Python, these algorithms have a potential drawback: some are not specifically designed for probabilistic estimation and could produce inaccurate probabilistic estimates. These probabilistic estimates, however, can be more reliable after model calibration; see the Appendix for further details. 

\subsection{Cross-Fitting}
\label{subsec:cross-fit}
Cross-fitting is one of the basic tools in machine learning for model evaluation and parameter selection. It has also received more and more attention in the field of statistics as a standard approach to creating valid confidence intervals. See, e.g., \citet{wager2016high, athey2016recursive, chernozhukov2018double}. In this paper, cross-fitting will serve to create pseudo external datasets when no external data sources are available.

In $K$-folds cross-fitting, the sample data are randomly split into $K$ subsets with equal sample size. In each loop, $(K-1)$ out of the $K$ data subsets are used as a pseudo external data for the estimation of the nuisance parameters ${\mu}_a$, and the remaining one is used for the interval estimation of $\theta$ based on the nuisance estimators. This fitting-estimation procedure repeats $K$ times, going through all leave-one-out subsets, and the final interval estimator of $\theta$ is the average of the $K$ interval estimators. The fitting-estimation procedure creates independency between the data for model training and the data for interval estimation, and thus justifies the non-parametric bounds estimators provided in section \ref{sec:nonpara-estimation}.

Cross-fitting itself can be taken as an action of partitioning. However, because the data splitting procedure is completely random, every split dataset is representative of the population. Therefore, the partitioning-based bounds under cross-fitting should not differ from the ordinary partitioning-based bounds. Even if certain covariate distribution shift occurs after data splitting due to the randomness involved in the splitting procedure, there is no need for concern as finer partitioning does not cause loss of information, as stated in Proposition \ref{prop:finerpartition}.

Although cross-fitting does not result in loss of information, it does affect the variances and convergence rates of the nonparametric bounds estimators provided in section \ref{sec:nonpara-estimation}. The number of folds, $K$, essentially yields a trade-off between the information (sharpness) of the bounds and the precisions of the bounds estimators. With more folds, on the one hand, we have more data in each loop for model training which leads to better estimators of the nuisance parameters and hence a more informative partition. On the other hand, the variances of the bounds estimators may get bigger as $K$ increases because of less data for bounds estimation in each loop. The $K$ can be determined according to the sample size. We suggest setting $K=2$ or $K=5$ in practice, and adopt $K=2$ in the numerical studies.

\section{Assessment of Estimated Bounds}
\label{sec:precision-evaluation}

One of the major issues for practitioners, after obtaining an interval estimator of $\theta$, is how to assess the precision of the interval estimator. Let us first focus on the case that external data are available and thus the partition $\Delta$ can be viewed as fixed (conditional on the external data). In this case, our estimated bounds $[\widehat{L}(\theta,\Delta), \widehat{U}(\theta,\Delta)]$ converge to $[{L}(\theta,\Delta), {U}(\theta,\Delta)]$ under randomization of the treatment assignments. The information on the precisions of the estimated bounds is included in the distribution of $\widehat{L}(\theta,\Delta)-{L}(\theta,\Delta)$ and $\widehat{U}(\theta,\Delta)-{U}(\theta,\Delta)$, of which the quantile estimators can be used to construct an extended confidence interval of $\theta$.

We first give a discussion on $\widehat{L}(\theta,\Delta)-{L}(\theta,\Delta)$. Provided a partition $\Delta=\{\mbX_1,\cdots,\mbX_J\}$, recall that
\[
\widehat{L}(\theta,\Delta) = \sum_{j=1}^Jn_j/n\max\{0,\hat{\mu}_{0}(\mbX_j)-\hat{\mu}_{1}(\mbX_j)\}.
\]
Let us proceed conditional on $\bn = (n_1,\cdots,n_J)$. Then,  the $\hat{\mu}_{a}(\mbX_j)\mid n_j$ will be asymptotically normally distributed as $n_j\rightarrow\infty$ with the mean $\tau_{a,j} = {\mu}_{a}(\mbX_j)$ and the variance $\sigma^2_{Y(a),j}/(n_j\pi_a)$ for $a=0,1$ under certain regularity conditions, and the variance of $\hat{\mu}_{0}(\mbX_j)-\hat{\mu}_{1}(\mbX_j)$ will be given by $\sigma^2_{Y(0),j}/(n_j\pi_0)+\sigma^2_{Y(1),j}/(n_j\pi_1)$ \citep{li2017general,ye2023toward}. Note that the mean $\tau_{a,j}$ does not rely on $n_j$. To be clear, we denote by $\bn^*=(n^*_1,\cdots,n^*_J)$ a random vector so that $\sum_{j=1}^Jn^*_j=n$ and $\bn$ is a realization of $\bn^*$. Then, the distribution of $\widehat{L}(\theta,\Delta) - {L}(\theta,\Delta)$ will be approximately $\sum_{j=1}^Jn^*_j/n\max\{0,N(\tau_{0,j}-\tau_{1,j}, {\sigma}_{1,j}^2+{\sigma}_{0,j}^2)\} - {L}(\theta,\Delta)$ where $\sigma^2_{a,j}=\sigma^2_{Y(a),j}/(n^*_j\pi_a)$ for $a=0,1$. Conditional on $\bn^*$, this distribution can be estimated as
\begin{align}
\label{eq: L-distribution}
\sum_{j=1}^Jn^*_j/n\max\{0,N(\hat{\mu}_{0}(\mbX_j)-\hat{\mu}_{1}(\mbX_j), \hat{\sigma}_{1,j}^2+\hat{\sigma}_{0,j}^2)\} - \widehat{L}(\theta,\Delta)\mid \mathcal{S},\bn^*
\end{align}
where $\hat{\sigma}_{a,j}^2={\sum_{i:\bX_i\in\mbX_j}1\{A_i=a\}(Y_i-\Bar{Y}_{a,j})^2}/\{n^*_{j}n_{a,j}/n_j(n^*_{j}n_{a,j}/n_j-1)\}$, $\Bar{Y}_{a,j}=n_{a,j}^{-1}\sum_{\bX_i\in\mbX_j}1\{A_i=a\}Y_i$, $n_{a,j}=\#\{i:A_i=a,\bX_i\in\mbX_j\}$ for $a=0,1$. Since $Y_i$ is binary, the sampling variance has a simple form: $\hat{\sigma}_{a,j}^2=n_{a,j}\Bar{Y}_{a,j}(1-\Bar{Y}_{a,j})/\{n^*_{j}n_{a,j}/n_j(n^*_{j}n_{a,j}/n_j-1)\}$. 
The distribution (\ref{eq: L-distribution}) can be simulated by Monte Carlo. Similarly, the distribution of $\widehat{U}(\theta,\Delta) - {U}(\theta,\Delta)\mid\bn$ can be approximated by
\begin{align}
\label{eq: U-distribution}
\sum_{j=1}^Jn^*_j/n\min\{N(\hat{\mu}_{0}(\mbX_j), \hat{\sigma}_{0,j}^2), 1 - N(\hat{\mu}_{1}(\mbX_j), \hat{\sigma}_{1,j}^2)\} - \widehat{U}(\theta,\Delta)\mid \mathcal{S},\bn^* .
\end{align}
The distribution of $\bn^*$ can be approximately by a multivariate hypergeometric distribution with sufficiently large $M$ (so that $n_jM/n$ are integers) as
\[
\pr(n^*_1=m_1,\cdots, n^*_J=m_J) = \left\{\begin{array}{cc}
   \frac{\prod_{j=1}^J\binom{n_jM/n}{m_j}}{\binom{M}{n}} & \text{if } \sum_{j=1}^Jm_j=n \\
    0 & o.w.
\end{array}\right..
\]
Based on the above results, we can numerically approximate the distribution $\sum_{j=1}^Jn^*_j/n\max\{0,N(\hat{\mu}_{0}(\mbX_j)-\hat{\mu}_{1}(\mbX_j), \hat{\sigma}_{1,j}^2+\hat{\sigma}_{0,j}^2)\} \mid \mathcal{S}$, which is an approximation of the distribution of $\widehat{L}(\theta,\Delta)$. Let $\hat{\xi}_{\alpha,L}$ be its $\alpha$-upper-quantile. Together with the constraint ${L}(\theta,\Delta)\in[0,1]$, the proposed two-sided $(1-\alpha)$ confidence interval for ${L}(\theta,\Delta)$ will be 
\[
C_{\alpha,L}(\Delta) = [\max\{2\widehat{L}(\theta,\Delta) -\hat{\xi}_{\alpha/2,L} ,0\},\min\{2\widehat{L}(\theta,\Delta) -\hat{\xi}_{1-\alpha/2,L},1\} ].
\]
Similarly, we can get a $(1-\alpha)$ confidence interval for ${U}(\theta,\Delta)$ as
\[
C_{\alpha,U}(\Delta) = [\max\{2\widehat{U}(\theta,\Delta) -\hat{\xi}_{\alpha/2,U},0\},\min\{2\widehat{U}(\theta,\Delta) -\hat{\xi}_{1-\alpha/2,U},1\} ]
\]
where $\hat{\xi}_{\alpha,U}$ is the estimated $\alpha$-upper-quantile of the distribution $\sum_{j=1}^Jn^*_j/n\min\{N(\hat{\mu}_{0}(\mbX_j), \hat{\sigma}_{0,j}^2), 1 - N(\hat{\mu}_{1}(\mbX_j), \hat{\sigma}_{1,j}^2)\}\mid \mathcal{S}$. However, since the proposed confidence intervals are constructed based on the assumption that $\hat{\mu}_{a}(\mbX_j)\mid n_j$ is approximately normal, they are valid only when $n_j,j=1,\cdots,J$ are large enough, and may fail if $n$ is small and $\rho_j=\pr(\bX\in\mbX_j)\approx 0$ for some $j$. After Bonferroni correction, an extended $(1-\alpha)$ confidence interval for $[L(\theta,\Delta),U(\theta,\Delta)]$ is given as
\[
C_{\alpha}(\theta,\Delta) = [\max\{2\widehat{L}(\theta,\Delta) -\hat{\xi}_{\alpha/2,L},0\},\min\{2\widehat{U}(\theta,\Delta) -\hat{\xi}_{1-\alpha/2,U},1\} ],
\]
which can be used as a $(1-\alpha)$ confidence interval for $\theta$. This extended confidence interval is, of course, too conservative for $\theta$ in general cases. It might over-cover $\theta$ even in the extreme case that $L(\theta,\Delta)=L(\theta)=\theta$ or $U(\theta,\Delta)=U(\theta)=\theta$ because of the Bonferroni correction. In the context of cross-fitting, the confidence intervals obtained from the split samples are aggregated by taking their average.

\section{Numerical Experiments}
\label{sec:numerical}

In this section, we evaluate the proposed partitioning-based interval estimators through Monte Carlo numerical experiments. We consider two partitions, $\Delta_0$ and $\Delta_{full}$, with the resulting interval estimators of $\theta$ respectively referred to as ``Naive'' and ``Oracle''. Also, we consider plug-in partitions induced by six probabilistic classification algorithms: Logistic regression (Logit), Naive Bayes (NBayes), Support Vector Machines (SVM, \cite{cortes1995support}) with Platt scaling \citep{platt1999probabilistic}, K-Nearest Neighbors (KNN), Random Forests (RF, \citet{breiman2001random}), and eXtreme Gradient Boosting (XGBoost, \citet{chen2016xgboost}). See the Appendix for a brief description and \citet{hastie2009elements} for further details of these classification algorithms. We adopt the Python package \texttt{scikit-learn} in our coding to realize the classification algorithms, and the folds of cross-fitting is set to $K=2$.

We can also estimate the sharp bounds of $\theta$ by directly plugging in the nuisance estimators in the bounds as $\widehat{L}(\theta)=\E[\max\{0,\hat{\mu}_{0,N}(\bX)-\hat{\mu}_{1,N}(\bX)\}\mid \mathcal{G}]$ and $\widehat{U}(\theta)=\E[\min\{\hat{\mu}_{0,N}(\bX),1-\hat{\mu}_{1,N}(\bX)\}\mid \mathcal{G}]$. Similarly, the cross-fitting strategy will be utilized in the estimation. We label the resulting estimators by ``Plug-in'' and include them in the simulation results for comparison. Furthermore, we calculate and present the cross-validation misclassification rates (MCR) for the classification algorithms applied to the Monte Carlo data.

The data generating procedure is as follows. The observed covariates are generated by $\bX=(X_1,\cdots,X_{10})^\intercal\equald N(0,\bI_{10})$. Let $\bZ=(3X_1^2, X_2X_3, X_3, X_4, X_5)^\intercal$. The potential outcome is generated as $Y(a)= 1\{\bbeta_{a}^\intercal\bZ + \beta_{a,0} + \varepsilon\}$ where $\varepsilon\equald N(0,\sigma_{\varepsilon}^2)$ and $\beta_{a,0}$ is chosen so that $\pr(Y(0)=1)=0.2$ and $\pr(Y(1)=1)=0.4$. The $\sigma_{\varepsilon}$ controls the magnitude of the noise factor and thus the information upper bound of the partitions. The treatment assignments are generated based on the completely randomized design under the treated ratio equal to $1/2$. We consider two scenarios for $\bbeta_{a}$:
\begin{itemize}
    \item \textbf{Scenario 1} (Homogeneous treatment effect). $\bbeta_a=(1,1,1,1,1)^\intercal$ for $a=0,1$.
    \item \textbf{Scenario 2} (Heterogeneous treatment effect). $\bbeta_0=(1,1,1,1,1)^\intercal$ and $\bbeta_1 = (-1.2,  1 , -0.8,  0.5, -0.3)^\intercal$.
\end{itemize}

\begin{table}[t!]
  \centering
  \caption{Monte Carlo simulation results for the interval estimators with $n=500$. All values except the percentages in the table are presented in units scaled by $10^{-2}$.}
  \resizebox{\linewidth}{!}{
    \begin{tabular}{lcclccrclccrcrr}
    \toprule
          &       &       & \multicolumn{4}{c}{Plug-in }   &       & \multicolumn{4}{c}{Partitioning} &       &       &  \\
\cmidrule{4-7}\cmidrule{9-12}    Method & $\theta$ &       & Estimate & Bias  & Width & CovR  &       & Estimate & Bias  & Width & CovR  &       & TMCR  & CMCR \\
    \midrule
    \multicolumn{15}{l}{Scenario 1, $\sigma_{\varepsilon} = 1$} \\
    Naive & 0.1   &       & $[0.0, 20.5]$ & 0.0   & 20.5  & 100.0\% &       & $[0.0, 20.5]$ & 0.0   & 20.5  & 100.0\% &       & NaN   & NaN \\
    KNN   & 0.1   &       & $[2.5, 11.7]$ & 2.4   & 9.2   & 0.0\% &       & $[0.1, 15.2]$ & 0.1   & 15.1  & 76.6\% &       & 31.3\% & 18.8\% \\
    RF    & 0.1   &       & $[0.8, 15.2]$ & 0.7   & 14.4  & 0.5\% &       & $[0.1, 5.3]$ & 0.1   & 5.2   & 76.5\% &       & 17.7\% & 10.6\% \\
    Oracle & 0.1   &       & NaN   & NaN   & NaN   & NaN   &       & $[0.0, 0.6]$ & 0.0   & 0.6   & 73.5\% &       & NaN   & NaN \\
    \midrule
    \multicolumn{15}{l}{Scenario 1, $\sigma_{\varepsilon} = 2$} \\
    Naive & 1.9   &       & $[0.0, 22.7]$ & 0.0   & 22.7  & 100.0\% &      & $[0.0, 22.7]$ & 0.0   & 22.7  & 100.0\% &       & NaN   & NaN \\
    KNN   & 1.9   &       & $[2.7, 13.8]$ & 0.9   & 11.1  & 21.2\% &       & $[0.2, 17.7]$ & 0.0   & 17.5  & 99.2\% &       & 35.1\% & 21.1\% \\
    RF    & 1.9   &       & $[0.9, 17.7]$ & 0.0   & 16.8  & 95.8\% &       & $[0.2, 8.6]$ & 0.0   & 8.4   & 100.0\% &       & 23.8\% & 14.1\% \\
    Oracle & 1.9   &       & NaN   & NaN   & NaN   & NaN   &       & $[0.0, 4.3]$ & 0.0   & 4.3   & 98.5\% &       & NaN   & NaN \\
    \midrule
    \multicolumn{15}{l}{Scenario 2, $\sigma_{\varepsilon} = 1$} \\
    Naive & 19.7  &       & $[0.0, 20.6]$ & 0.6   & 20.6  & 61.6\% &       & $[0.0, 20.6]$ & 0.6   & 20.6  & 61.6\% &       & NaN   & NaN \\
    KNN   & 19.7  &       & $[3.0, 12.4]$ & 7.2   & 9.5   & 0.1\% &       & $[7.3, 20.4]$ & 0.6   & 13.2  & 62.3\% &       & 35.9\% & 18.7\% \\
    RF    & 19.7  &       & $[7.6, 20.5]$ & 0.5   & 12.9  & 64.2\% &       & $[16.7, 20.4]$ & 0.7   & 3.7   & 49.5\% &       & 21.7\% & 10.6\% \\
    Oracle & 19.7  &       & NaN   & NaN   & NaN   & NaN   &       & $[19.2, 19.8]$ & 1.4   & 0.6   & 11.4\% &       & NaN   & NaN \\
    \midrule
    \multicolumn{15}{l}{Scenario 2, $\sigma_{\varepsilon} = 2$} \\
    Naive & 20.6  &       & $[0.0, 22.6]$ & 0.3   & 22.6  & 75.9\% &       & $[0.0, 22.6]$ & 0.3   & 22.6  & 75.9\% &       & NaN   & NaN \\
    KNN   & 20.6  &       & $[3.5, 14.6]$ & 6.1   & 11.1  & 1.3\% &       & $[6.2, 22.4]$ & 0.3   & 16.2  & 74.7\% &       & 38.5\% & 21.0\% \\
    RF    & 20.6  &       & $[7.4, 22.5]$ & 0.3   & 15.1  & 78.5\% &       & $[15.6, 22.3]$ & 0.3   & 6.7   & 73.7\% &       & 28.6\% & 14.0\% \\
    Oracle & 20.6  &       & NaN   & NaN   & NaN   & NaN   &       & $[18.6, 21.8]$ & 0.6   & 3.3   & 51.9\% &       & NaN   & NaN \\    
    \bottomrule
    \end{tabular}%
    }
  \label{tab:estimation}%
\end{table}%

For each setup, we generate $M=1000$ Monte Carlo datasets. Let $\widehat{B}(\theta) = [\widehat{L}(\theta),\widehat{U}(\theta)]$ denote an interval estimator of $\theta$ and $\widehat{B}_m(\theta) = [\widehat{L}_m(\theta),\widehat{U}_m(\theta)]$ an interval estimate based on the $m$-th Monte Carlo dataset. We adopt the following three metrics to evaluate $\widehat{B}(\theta)$ based on the $M$ Monte Carlo replications:
\begin{itemize}
    \item \textbf{Bias}: $\text{Bias}(\widehat{B}(\theta)) = M^{-1}\sum_{m=1}^M\max\{\widehat{L}_m(\theta)-\theta,\theta-\widehat{U}_m(\theta),0\}$.
    \item \textbf{Width}: $\text{Width}(\widehat{B}(\theta)) = M^{-1}\sum_{m=1}^M(\widehat{U}_m(\theta)-\widehat{L}_m(\theta))$.
    \item \textbf{Coverage rate (CovR)}: $\text{CovR}(\widehat{B}(\theta)) = M^{-1}\sum_{m=1}^M1\{\widehat{L}_m(\theta)\leq\theta\leq \widehat{U}_m(\theta)\}$.
\end{itemize}

\begin{figure}[t!]
    \centering
    \begin{subfigure}[b]{0.45\textwidth}
        \centering
        \includegraphics[width=\textwidth]{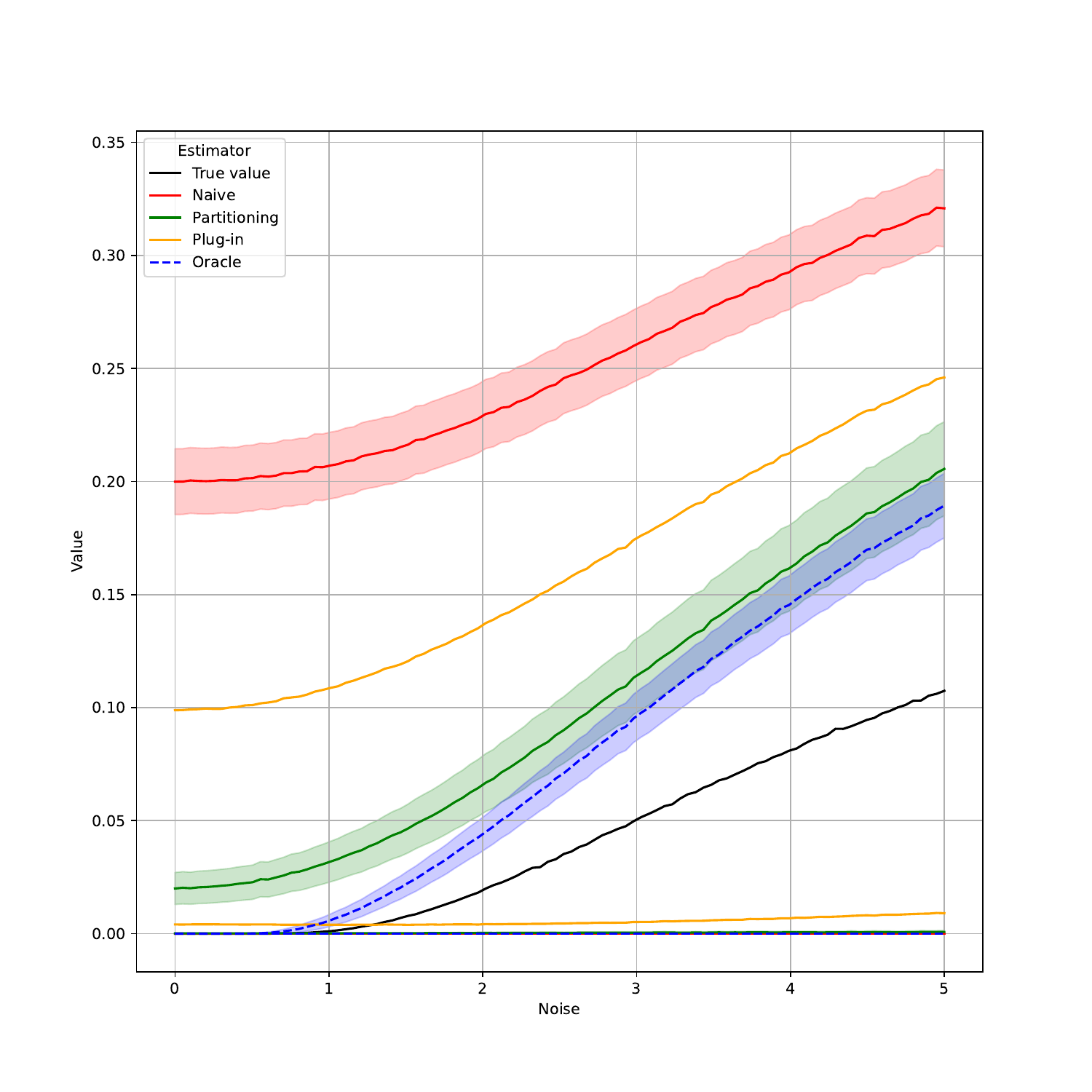} 
    \end{subfigure}
    \begin{subfigure}[b]{0.45\textwidth}
        \centering
        \includegraphics[width=\textwidth]{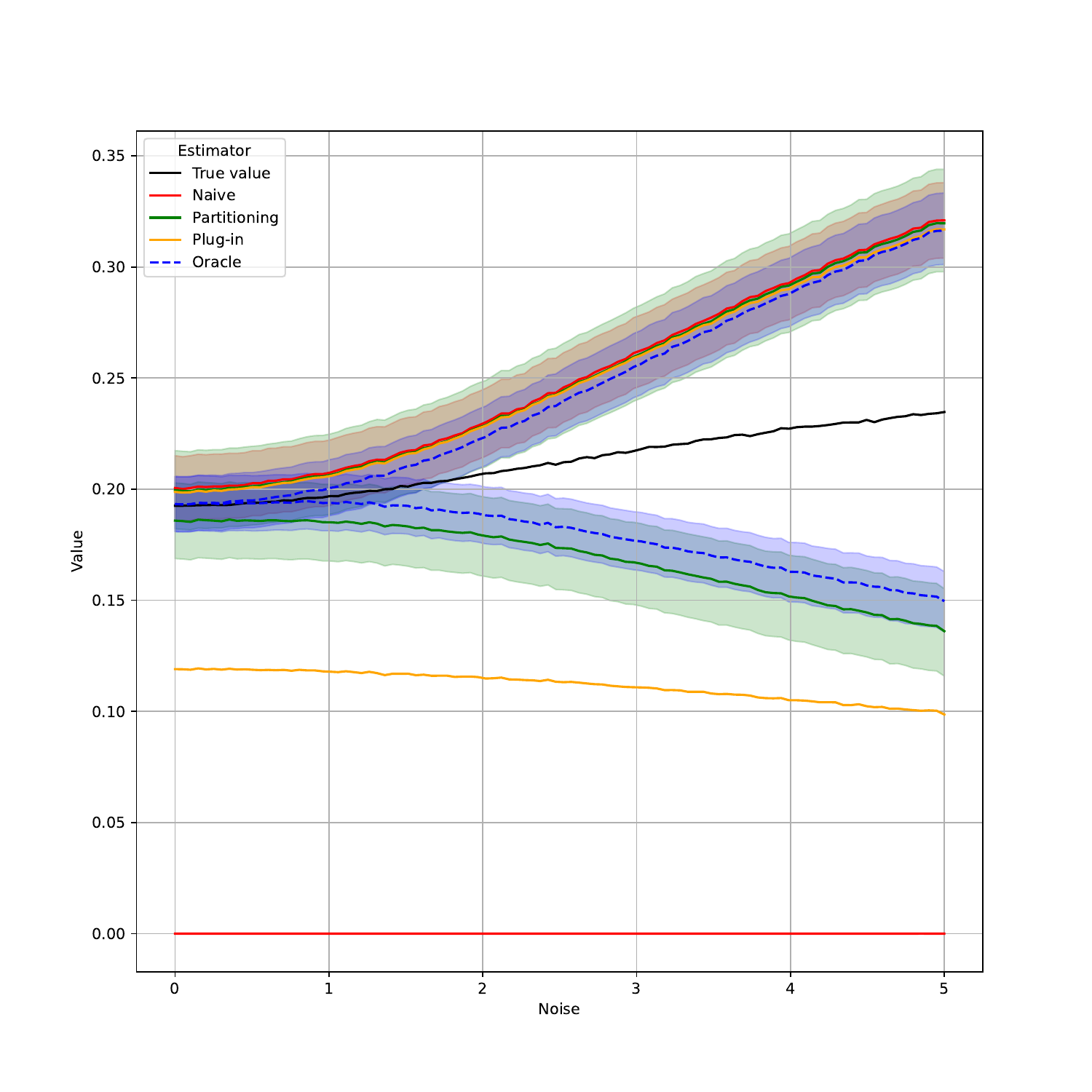} 
    \end{subfigure}
    \caption{Naive and Oracle bounds, estimated plug-in and partitioning-based bounds using the RF algorithm, and their corresponding $75\%$ confidence bands. X-axis denotes the value of $\sigma_{\varepsilon}$. The plot for Scenario 1 is presented in the left panel, and the plot for Scenario 2 is presented in the right panel.}
    \label{fig:RF}
\end{figure}

\begin{figure}[t!]
    \centering
    \begin{subfigure}[b]{0.45\textwidth}
        \centering
        \includegraphics[width=\textwidth]{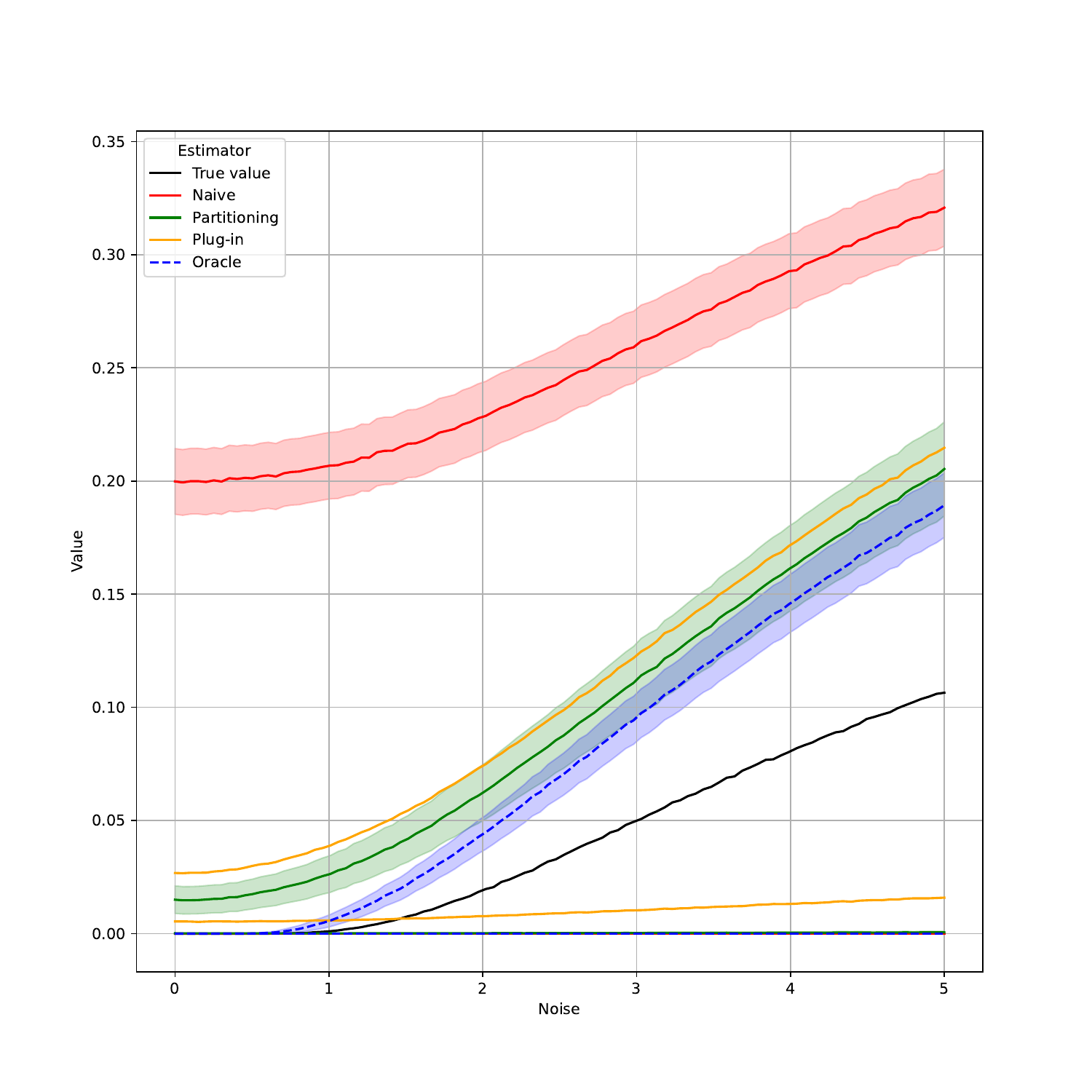} 
    \end{subfigure}
    \begin{subfigure}[b]{0.45\textwidth}
        \centering
        \includegraphics[width=\textwidth]{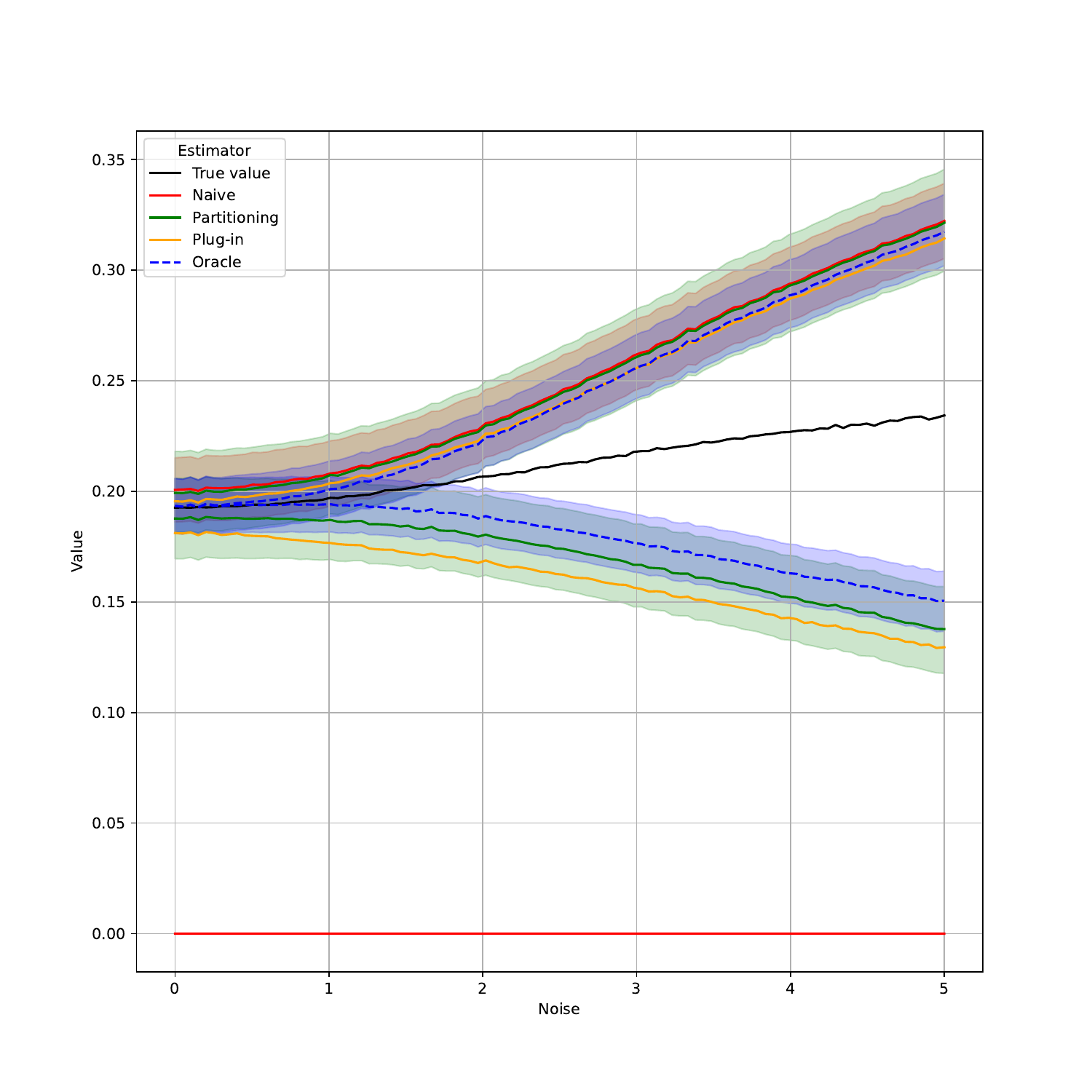} 
    \end{subfigure}
    \caption{Naive and Oracle bounds, estimated plug-in and partitioning-based bounds using the RF algorithm combined with model calibration, and their corresponding $75\%$ confidence bands. X-axis denotes the value of $\sigma_{\varepsilon}$. The plot for Scenario 1 is presented in the left panel, and the plot for Scenario 2 is presented in the right panel.}
    \label{fig:RF-calib}
\end{figure}

Let $n$ denote the size of the sample data in each Monte Carlo replication. We consider combinations of the parameters $\sigma_{\varepsilon}\in\{1,2\}$ and $n\in\{500,2000\}$. Simulation results for KNN and RF with $n=500$ are summarized in Table \ref{tab:estimation}. Additional results for  the six aforementioned classification algorithms, with model-calibration, and for $n=2000$, are given in the Appendix. The metrics TMCR and CMCR in the table denotes respectively the empirical cross-validation misclassification rate of the corresponding classification algorithm on the treated and the control group data. Results presented in Table \ref{tab:estimation} can be summarized as follows: (1) the proposed partitioning-based interval estimators have much smaller biases as compared to the plug-in based interval estimators; (2) under the RF algorithm, the partitioning-based interval estimators are more informative than the plug-in based estimators with smaller widths; under the KNN algorithm, the plug-in based estimators are narrower, but they are severely biased and fail to cover the true value. For illustrations of a more complete picture, we vary the value of $\sigma_{\varepsilon}$ from $0$ to $5$ and conduct simulations for the interval estimators under the RF algorithm with $n=2000$. The estimated upper and lower bounds and 75\% confidence bands are depicted in Figure \ref{fig:RF}. We observe that the partitioning-based estimated bounds are much more informative than the plug-in based estimated bounds and close to the Oracle bounds. The results under model calibration are given in Figure \ref{fig:RF-calib}, where the plug-in based estimated bounds perform much better in terms of bias because of the probabilities correction, despite they are still wider than the partitioning-based interval estimators. However, when $\sigma_{\varepsilon}$ is small the plug-in based estimated lower bound is biased and greater than the true value of interest.  

Additional tables for other simulation settings presented in the Appendix demonstrate similar patterns. A brief summary on extended confidence intervals: a significance level of $\alpha=0.25$ leads to around $98\%$ coverage rates of the extended confidence intervals on $\theta$ under all the classification algorithms in our setups. Additional figures on the results under the KNN algorithm and the tables on evaluation of the confidence intervals developed in section \ref{sec:precision-evaluation} are also provided in the Appendix. In summary, the empirical coverage rates of the proposed confidence intervals are very close to the nominal levels, except some extreme cases as explained in the Appendix.

\section{Analysis of the AIDS Clinical Trials Group Study 175 Data}
\label{sec:real-data}

AIDS Clinical Trials Group Study 175 (ACTG 175) is a randomized clinical trial that aimed to compare the treatment effects of different therapies on patients infected with human immunodeficiency virus type 1 (HIV-1) whose CD4 cell counts were between 200 and 500 per cubic millimeter \citep{hammer1996trial}. Four therapy treatments to the clinical subjects were considered and randomized in the study: monotherapy with zidovudine alone, monotherapy with didanosine alone, combination therapy with zidovudine and didanosine, and combination therapy with zidovudine and zalcitabine. In the public ACTG 175 dataset that is available online, the CD4 cell counts of all the patients were recorded at baseline prior to the treatments and 20+/-5 weeks after the treatments. The CD4 cell counts reveal the damage of HIV-1 infection to the immune system, as which progresses the CD4 cell counts decrease. We define a new binary outcome as the indicator of whether the CD4 cell counts had increased at 20+/-5 weeks since the baseline, denoted by $Y$, to assess the treatment effects of different therapies, with $Y=1$ being favorable. We focus on the comparison between the combination therapy with zidovudine and zalcitabine, termed as the treated group ($A=1$), and monotherapy with zidovudine alone, termed as the control group ($A=0$). The average treatment effect is estimated as $0.119$ with the $95\%$ confidence interval as $[0.059,0.179]$, implying the combination therapy is, on average, beneficial to the clinical subjects during the first 20 weeks. 

We are interested in the question ``What is the potential proportion of patients getting physically worse after the combination therapy''? To answer this question, we apply our proposed partitioning-based interval estimation approach to estimate the treatment harm rate. The interval estimates and the 75\% confidence intervals for the upper and lower bounds as well as the extended confidence intervals for the THR are given in Table \ref{tab:real-data}. The naive Fr\'echet-Hoeffding bounds are $[0, 0.436]$, implying no more than $43.6\%$ patients might be negatively affected by the combination therapy. The partitioning-based interval estimators are more precise after covariate adjustment and substantially sharpen the upper bound. For example, the partitioning-based interval estimate of the THR under the RF algorithm and two-folds cross-fitting is $[0.004, 0.318]$, indicating that a potential of $0.4\%$ treatment harm rate but no greater than $31.8\%$. If we take the possible variability into consideration, a $75\%$ extended confidence interval of the THR under the RF is given as $[0, 0.353]$, indicating that there is no evidence that the combination therapy does harm to the patients.

\begin{table}[t!]
  \centering
  \caption{Interval estimates of the treatment harm rate in ACTG 175 and $75\%$ confidence intervals for the partitioning-based upper and lower bounds.}
  \resizebox{\textwidth}{!}{
    \begin{tabular}{lllllrllll}
    \toprule
          & \multicolumn{4}{c}{Two folds} &       & \multicolumn{4}{c}{Five folds} \\
\cmidrule{2-5}\cmidrule{7-10}     Method   & Estimate & Upper-CI & Lower-CI & Extended-CI &       & Estimate & Upper-CI & Lower-CI & Extended-CI \\
    \midrule
    Naive & $[0, 0.436]$ & $[0.424, 0.465]$ & $[0, 0]$ & $[0, 0.465]$ &       & $[0, 0.436]$ & $[0.424, 0.465]$ & $[0, 0]$ & $[0, 0.465]$ \\
    Logit & $[0.003, 0.34]$ & $[0.309, 0.375]$ & $[0, 0.006]$ & $[0, 0.375]$ &       & $[0.008, 0.333]$ & $[0.296, 0.397]$ & $[0, 0.015]$ & $[0, 0.397]$ \\
    NBayes & $[0.002, 0.355]$ & $[0.326, 0.391]$ & $[0, 0.004]$ & $[0, 0.391]$ &       & $[0.006, 0.348]$ & $[0.309, 0.419]$ & $[0, 0.009]$ & $[0, 0.419]$ \\
    SVM   & $[0, 0.329]$ & $[0.297, 0.364]$ & $[0, 0]$ & $[0, 0.364]$ &       & $[0.015, 0.317]$ & $[0.271, 0.375]$ & $[0, 0.03]$ & $[0, 0.375]$ \\
    KNN   & $[0, 0.356]$ & $[0.323, 0.391]$ & $[0, 0]$ & $[0, 0.391]$ &       & $[0.011, 0.346]$ & $[0.308, 0.409]$ & $[0, 0.018]$ & $[0, 0.409]$ \\
    RF    & $[0.004, 0.318]$ & $[0.285, 0.353]$ & $[0, 0.009]$ & $[0, 0.353]$ &       & $[0.011, 0.344]$ & $[0.304, 0.406]$ & $[0, 0.019]$ & $[0, 0.406]$ \\
    XGBoost & $[0.001, 0.358]$ & $[0.329, 0.395]$ & $[0, 0.001]$ & $[0, 0.395]$ &       & $[0.016, 0.344]$ & $[0.303, 0.408]$ & $[0, 0.027]$ & $[0, 0.408]$ \\
    \bottomrule
    \end{tabular}%
    }
  \label{tab:real-data}%
\end{table}%

\section{Concluding Remarks}
\label{CR}

In this paper, we developed a class of partitioning-based bounds for the THR which can be consistently estimated with RCT data. Our proposed methodologies do not impose any untestable identification or ambiguous model assumptions for the estimation procedures other than the randomization which is part of the design. Our interval estimators not only achieve information gains from the baseline auxiliary covariates but also remain valid under model-misspecification, which is a highly desirable model-assisted property.

The idea of partitioning can be viewed as a generalization of stratification, where some covariates are discrete or categorical, and the covariates space can be naturally partitioned based on the levels of these covariates. In the context of estimating the average treatment effect, stratification serves as a covariate adjustment method to improve the estimator's efficiency. In the context of estimating the THR, the idea of using stratification to improve the informativeness of the bounds has been previously mentioned by, for example, \citet{zhang2013assessing} and \citet{huang2017inequality}. However, there are no guarantees in adequate information gains after stratification relying on the given levels of the categorical covariates. If there are a large number of levels within the covariates, the size of the strata may become excessively large. The proposed partitioning-based method provides a systematic approach to achieving an informative partition with a maximum of four cells and is applicable to data containing any types of covariates. The modern machine learning methods can be easily incorporated into our framework to further improve the interval estimators as discussed in our paper.

One possible limitation of our proposed methods is that they are specifically designed for binary outcomes. They cannot be directly applied to continuous or ordinal outcomes. Also, we only consider simple randomization designs such as completely randomized trials and Bernoulli trials. It might be of interest to practitioners to consider blocking randomization and covariate-adaptive randomization which are often used in practice. These extensions will be pursued in future research.

\section*{Software and Data}

The partitioning-based algorithm for treatment effect heterogeneity has been developed into a Python package named \texttt{partbte} avaiable on \href{https://github.com/w62liang/partition-te}{https://github.com/w62liang/partition-te}. It relies on the estimators in the Python package \texttt{scikit-learn} \citep{pedregosa2011scikit} for nuisance estimation and can be used to estimate $\pr(Y(1)=s,Y(0)=t)$ for $s,t\in\{-1,1\}$. The ACTG 175 dataset is publicly available in the R package \texttt{BART} \citep{sparapani2021nonparametric}.

\section*{Acknowledgment}

This research was supported by a grant to C. Wu from the Natural Sciences and Engineering Research Council of Canada. 

\section*{Appendix}

\subsection*{A1. Proofs of Main Propositions}

\medskip\noindent
{\it \bf Proof of Proposition \ref{prop:fullpartition}}
\medskip\noindent

\begin{proof}
Let $\rho_{a,t}=\pr(\bX\in\mbX_{a,t})$ for $a=0,1$, $t=-1,1$. It follows that
\begin{align*}
    I_{\theta}(\Delta_{full})=&\sum_{(a,t)\in\mbL}\rho_{a,t}\max\{\mu_{1}(\mbX_{a,t}),1-\mu_{1}(\mbX_{a,t}),\mu_{0}(\mbX_{a,t}),1-\mu_{0}(\mbX_{a,t})\}\\
    \geq & \sum_{(a,t)\in\mbL}\rho_{a,t}\{(1-t)/2+t\mu_{a}(\mbX_{a,t})\}\\
    \geq & \sum_{(a,t)\in\mbL}\int_{\mbX_{a,t}} \mu_{max}(\bx)f(\bx)v(d\bx)\\
     \geq& \int_{\mbX} \mu_{max}(\bx)f(\bx)v(d\bx) = I_{\theta}(\Delta_{\infty}).
\end{align*}
Since $I_{\theta}(\Delta_{full})$ reaches the information upper bound, we must have $B_{\theta}(\Delta_{full}) = B_{\theta}(\Delta_{\infty})$.
\end{proof}

\medskip\noindent
{\it \bf Proof of Proposition \ref{prop:partition-bayes}}
\medskip\noindent

\begin{proof}
Some algebra shows that 
\begin{align*}
m_1(g) + m_0(g) &= \pr(g(\bX)\neq (1,Y(1))) + \pr(g(\bX)\neq (0,Y(0))) \\
& = \E\left\{\sum_{(a,t)\in\mbL}\pr(Y(a)=t\mid\bX)1\{g(\bX)\neq (a,t)\} \right\}\\
& = C -\E\left\{\sum_{(a,t)\in\mbL}\pr(Y(a)=t\mid\bX)1\{g(\bX)= (a,t)\} \right\}
\end{align*}
for some constant $C$. This loss is minimized at $g^*(\bx) = \underset{(a,t)\in\mbL}{\operatorname{argmax}}\pr(Y(a)=t\mid\bX=\bx)$.
\end{proof}

\medskip\noindent
{\it \bf Proof of Proposition \ref{prop:nonpara-est}}
\medskip\noindent

\begin{proof}
Let $\delta_{i,j}=1\{\bX_i\in\mbX_j\}$, $\mathcal{F}_j=\{\delta_{i,j},i=1,\cdots,n\}$, and $\mathcal{A}=\{A_1,\cdots,A_n\}$. It is easy to see that $\E[\hat{\mu}_{a}(\mbX_j)|\mathcal{A},\mathcal{F}_j]={\mu}_a(\mbX_j)$ for $a=0,1$. For any $\varepsilon>0$, the Chebyshev's inequality gives
\begin{align*}
    \pr(|\hat{\mu}_{a}(\mbX_j)-{\mu}_a(\mbX_j)|\geq \varepsilon|\mathcal{A},\mathcal{F}_j)\leq \varepsilon^{-2}\var(\hat{\mu}_{a}(\mbX_j)|\mathcal{A},\mathcal{F}_j)=\frac{\sigma^2_{Y(a),j}}{\varepsilon^{2}\sum_{i=1}^n1\{A_i=a\}\delta_{i,j}}
\end{align*}
where $\sigma^2_{Y(a),j}=\var(Y(a)|\bX\in\mbX_j)$. Then, we have
\begin{align*}
    \pr(|\hat{\mu}_{a}(\mbX_j)-{\mu}_a(\mbX_j)|\geq \varepsilon)=\E_{\mathcal{A},\mathcal{F}_j}\pr(|\hat{\mu}_{a}(\mbX_j)-{\mu}_a(\mbX_j)|\geq \varepsilon|\mathcal{A},\mathcal{F}_j)\leq \frac{\sigma^2_{Y(a),j}}{\varepsilon^{2}n\pi_a\rho_j}
\end{align*}
and thus $\hat{\mu}_{a}(\mbX_j)={\mu}_a(\mbX_j)+O_p(\sigma_{Y(a),j}/\sqrt{n\pi_a\rho_j})$. Some algebra shows
\begin{align*}
    \widehat{L}(\theta,\Delta) &= \sum_{j=1}^J\frac{n_j}{n}\max\{0,\hat{\mu}_{0}(\mbX_j)-\hat{\mu}_{1}(\mbX_j)\}\\
    & = \sum_{j=1}^J(\frac{n_j}{n}-\rho_j)\max\{0,\hat{\mu}_{0}(\mbX_j)-\hat{\mu}_{1}(\mbX_j)\} + \sum_{j=1}^J\rho_j\max\{0,\hat{\mu}_{0}(\mbX_j)-\hat{\mu}_{1}(\mbX_j)\}.
\end{align*}
The first term at the right side of the above equation is $O_p(1/\sqrt{n})$ because ${n_j}/{n}=\rho_j+O_p(1/\sqrt{n})$ by the Markov inequality and $\max\{0,\hat{\mu}_{0}(\mbX_j)-\hat{\mu}_{1}(\mbX_j)\}\leq 1$. For the second term,
\begin{align*}
    \sum_{j=1}^J\rho_j\max\{0,\hat{\mu}_{0}(\mbX_j)-\hat{\mu}_{1}(\mbX_j)\} &\leq \sum_{j=1}^J\rho_j\max\{0,\hat{\mu}_{0}(\mbX_j)-{\mu}_0(\mbX_j)-\hat{\mu}_{1}(\mbX_j)+{\mu}_1(\mbX_j)\} + {L}(\theta,\Delta)\\
    &= {L}(\theta,\Delta)+O_p\left(\sum_{j=1}^J\sum_{a=0}^1\sigma_{Y(a),j}\sqrt{{\rho_j}/{{n\pi_a}}}\right).
\end{align*}
Similarly, for the estimated upper bound, we have
\begin{align*}
    \widehat{U}(\theta,\Delta) &= \sum_{j=1}^J\frac{n_j}{n}\min\{\hat{\mu}_{0}(\mbX_j),1-\hat{\mu}_{1}(\mbX_j)\}\\
    & = \sum_{j=1}^J(\frac{n_j}{n}-\rho_j)\min\{\hat{\mu}_{0}(\mbX_j),1-\hat{\mu}_{1}(\mbX_j)\} + \sum_{j=1}^J\rho_j\max\{\hat{\mu}_{0}(\mbX_j),1-\hat{\mu}_{1}(\mbX_j)\}\\
    & \leq \sum_{j=1}^J\rho_j\max\{{\mu}_0(\mbX_j),1-{\mu}_1(\mbX_j)\} + |\hat{\mu}_{0}(\mbX_j) - {\mu}_0(\mbX_j)| + |{\mu}_1(\mbX_j) - \hat{\mu}_{1}(\mbX_j)| + O_p(1/\sqrt{n})\\
    & = U(\theta,\Delta) + O_p\left(\sum_{j=1}^J\sum_{a=0}^1\sigma_{Y(a),j}\sqrt{{\rho_j}/{{n\pi_a}}} + 1/\sqrt{n}\right).
\end{align*}
For the expected information of the estimated bounds, some algebra shows
\begin{align*}
    \E[\hat{I}_{\theta}(\Delta)|\bn = \{n_1,\cdots,n_J\}] &= \sum_{j=1}^J\frac{n_j}{n}\E[\max\{\hat{\mu}_{0}(\mbX_j),1-\hat{\mu}_{0}(\mbX_j),\hat{\mu}_{1}(\mbX_j),1-\hat{\mu}_{1}(\mbX_j)\}|\bn]\\
    & \geq \sum_{j=1}^J\frac{n_j}{n}\max\{\E[\hat{\mu}_{0}(\mbX_j)|\bn],1-\E[\hat{\mu}_{0}(\mbX_j)|\bn],\E[\hat{\mu}_{1}(\mbX_j)|\bn],1-\E[\hat{\mu}_{0}(\mbX_j)|\bn]\}\\
    & \geq \sum_{j=1}^J\frac{n_j}{n}\max\{{\mu}_0(\mbX_j),1-{\mu}_0(\mbX_j),{\mu}_1(\mbX_j),1-{\mu}_1(\mbX_j)\}.
\end{align*}
The second inequality is because of the Jensen's inequality. It follows that $\E[\hat{I}_{\theta}(\Delta)]=\E\E[\hat{I}_{\theta}(\Delta)|\bn]\geq \sum_{j=1}^J\rho_j\max\{{\mu}_0(\mbX_j),1-{\mu}_0(\mbX_j),{\mu}_1(\mbX_j),1-{\mu}_1(\mbX_j)\}={I}_{\theta}(\Delta)$.
\end{proof}

\medskip\noindent
{\it \bf Proof of Proposition \ref{prop: plug-in partition}}
\medskip\noindent

\begin{proof}
Consider the information carried by $\Delta_{plug}$,  we have 
\begin{align*}
    I_{\theta}(\Delta_{plug}) &= \sum_{(a,t)\in\mbL}\rho_{a,t}\max\{\mu_1(\hat{\mbX}_{a,t}),1-\mu_1(\hat{\mbX}_{a,t}),\mu_0(\hat{\mbX}_{a,t}),1-\mu_0(\hat{\mbX}_{a,t})\}\\
    &\geq \sum_{(a,t)\in\mbL}\rho_{a,t} \{(1-t)/2+t\mu_a(\hat{\mbX}_{a,t})\}\\
    &\geq \sum_{(a,t)\in\mbL}\int_{\hat{\mbX}_{a,t}}\{(1-t)/2+t\mu_a(\bx)\}f_{\bX}(\bx)\nu(d\bx)\\
    &\geq \int_{\mbX} \mu_{max}(\bx)f_{\bX}(\bx)\nu(d\bx) + \sum_{(a,t)\in\mbL}\int_{\hat{\mbX}_{a,t}}\{(1-t)/2+t\mu_a(\bx)-\mu_{max}(\bx)\}f_{\bX}(\bx)\nu(d\bx)\\
    &\geq I_{\theta}(\Delta_{\infty}) + \sum_{(a,t)\in\mbL}\int_{\hat{\mbX}_{a,t}}\{t(\mu_a(\bx) - \hat{\mu}_{a}(\bx))+\hat{\mu}_{max,N}(\bx)-\mu_{max}(\bx)\}f_{\bX}(\bx)\nu(d\bx)\\
    &\geq I_{\theta}(\Delta_{\infty}) - \int_{{\mbX}}\{(\mu_{max}(\bx) - \hat{\mu}_{max,N}(\bx)) +\max_{a=0,1}|\mu_a(\bx) - \hat{\mu}_{a}(\bx)|\}f_{\bX}(\bx)\nu(d\bx).
\end{align*}  
\end{proof}

\subsection*{A2. Model Calibration and Isotonic Regression}

Model calibration in classification problems is a separate topic of long history and great importance in machine learning \citep{guo2017calibration,vaicenavicius2019evaluating,minderer2021revisiting}. It aims to provide more trustworthy probabilistic estimates while not changing the ranking performance of the original classification algorithms. Suppose we want to estimate $\mu(\bx)=\pr(Y=1|\bX=\bx)$ with the sample data $\{(Y_i,\bX_i),i=1,\cdots,n\}$ as $n$ i.i.d. copies of $(Y,\bX)$ where $Y$ is a binary response variable and $\bX$ is a vector of covariates. In our context of causal inference, the sample data will be the treated observations $\{(Y_i,\bX_i),A_i=1\}$ if we want to estimate $\mu_1(\bx)=\pr(Y=1|A=1,\bX=\bx)$ and the control observations $\{(Y_i,\bX_i),A_i=0\}$ if we want to estimate $\mu_0(\bx)=\pr(Y=1\mid A=0,\bX=\bx)$. Let $\hat{\mu}$ be an estimator of $\mu$ at hand, probably learned from the sample data. We focus in this section of the Appendix on why the model calibration can bring us better estimators of $\mu$ and how to construct a well-calibrated estimator based on $\hat{\mu}$.

\medskip\noindent
{\bf Definition A.1 (Well-calibrated estimator)}\,\, {\it  An estimator $\hat{\mu}$ is said well-calibrated if $\pr(Y(a)=1\mid\hat{\mu}(\bX))=\hat{\mu}(\bX)$ over the support of $\bX$.}
\medskip\noindent

\medskip\noindent
{\bf Proposition A.1}\,\, {\it The true model $\mu$ is well-calibrated.}
\medskip\noindent

It is often difficult and sometimes even impossible to get a perfect estimator of $\mu$. However, the true model is necessary to be well-calibrated. We can therefore reasonably speculate that an uncalibrated estimator will be made closer to the truth after calibration, and shift our attention to seeking a better-calibrated probabilistic classifier. Indeed, in our numerical experiments, the interval estimators of $\theta$ perform better after we calibrate the nuisance estimators. 

\medskip\noindent
{\bf Proposition A.2}\,\, {\it Let $g(z)=\pr(Y(a)=1\mid\hat{\mu}(\bX)=z)$ for $z\in[0,1]$. Then, $g\circ\hat{\mu}$ is well-calibrated.}
\medskip\noindent

Proposition A.2 provides a way to calibrate an estimator: the key is to find the conditional probability $g(z)=\pr(Y(a)=1\mid\hat{\mu}(\bX)=z)$, which refers to a calibrator. This may look similar to an ordinary binary regression problem, but they are different in the following three aspects. First, the input of the calibrator are probabilistic estimates in the range $[0,1]$, thus the regression is expected to be done in a unit square. Second, in order to retain the properties of the algorithms on classification, we would not like to change the rankings of the estimated probabilities after calibration, which implies that $g(z)$ should be a monotonically increasing function. Third,  if $\hat{\mu}$ is obtained based on the sample data, there might be over-fitting issues if we use the sample data to estimate the calibrator as well. The last concern, luckily, can be easily addressed by a popular technique in machine learning called cross-fitting.

Platt scaling \citep{platt1999probabilistic} and isotonic regression \citep{ayer1955empirical} are two commonly used approaches in model calibration for the estimation of a calibrator \citep{niculescu2005predicting}. Platt scaling is a parametric method that assumes a logistic model for $g(z)$. The isotonic regression, also known as monotone regression, is non-parametric. We focus on isotonic regression because it is more flexible and generally performs better than Platt scaling in practice.

Without loss of generality, we assume that $\hat{\mu}(\bX_1)\leq \hat{\mu}(\bX_2)\leq \cdots \leq \hat{\mu}(\bX_n)$. The isotonic regression assumes $g(z)$ is a step function as $g(z)=n^{-1}\sum_{i=1}^n 1\{z\leq w_i\}$ for some $0\leq w_1,\cdots,w_n \leq 1$ and estimates $w_1,\cdots,w_n$ by maximizing $\sum_{i=1}^n(Y_i-w_i)^2$ with respect to the constraints $0\leq w_1\leq w_2\leq\cdots\leq w_n\leq 1$. It is clear that the sequence $\hat{\mu}(\bX_i),i=1,\cdots,n$ maintains the order after isotonic regression because $g(z)$ is monotonically increasing. This quadratic optimization problem can be solved by an algorithm called pool-adjacent-violators (PAV) algorithm \citep{ayer1955empirical,busing2022monotone}. The cross-fitting isotonic regression algorithm can be rapidly implemented by a function named \texttt{CalibratedClassifierCV} in the Python library \texttt{scikit-learn}. 

\subsection*{A3. A Brief Review of Probabilistic Classification Algorithms}

We give a brief review of the six classification algorithms adopted in our numerical experiments and on how they produce the probabilistic estimates. Similarly, we assume that we want to estimate $\mu(\bx)=\pr(Y=1\mid\bX=\bx)$ with the sample data $\{(Y_i,\bX_i),i=1,\cdots,n\}$ as $n$ i.i.d. copies of $(Y,\bX)$ where $Y$ is a binary response variable and $\bX$ is a vector of covariates.

\medskip\noindent
{\bf Logistic Regression.} {\it In logistic regression, a parametric model $f(\bX^\intercal\bm{\beta})=1/(1+\exp(-\bX^\intercal\bbeta))$ is postulated for $\mu(\bX)$ and an estimator $\hat{\bbeta}$ of $\bbeta$ is computed by maximizing the regularized log-likelihood function
\[
\frac{1}{n}\sum_{i=1}^n\{Y_i\log f(\bX_i^\intercal\bm{\beta}) + (1-Y_i)\log(1-f(\bX_i^\intercal\bm{\beta}))\} + \lambda_n ||\bbeta||.
\]
The conditional probability is estimated as $\hat{\mu}(\bX)=f(\bX^\intercal\hat{\bbeta})$.
}
\medskip\noindent

\medskip\noindent
{\bf Gaussian Naive Bayes.} {\it In the perspective of Bayesian statistics, the conditional probability $\mu(\bx)$ is $\mu(\bx)=P(Y=1)f_1(\bx)/(P(Y=1)f_1(\bx)+P(Y=-1)f_0(\bx))$ where $f_t(\bx)$ is the density of $\bX$ given $Y=t$ for $t=-1,1$. For the $i$-th individual, let $\bX_i=(X_{i,1},\cdots,X_{i,p})^\intercal$. In Gaussian Naive Bayes, it is assumed that the covariates are independent of each other so that $f_t(\bX_i)=\prod_{k=1}^pf_{t,k}(\bX_{i,k})$ where $f_{t,k}(\bX_{i,k})$ is the density of $\bX_{i,k}$ given $Y=t$, and $f_{t,k}(\bX_{i,k})$ is modeled by the standard normal density function. For more general cases, one can consider one-dimensional kernel density estimates for the margins. 
}
\medskip\noindent

\medskip\noindent
{\bf K-Nearest-Neighbors.} {\it The K-nearest-neighbors (KNN) algorithm is very simple and does not require any parametric assumptions on the form of $\mu(\bX)$. The simplest KNN estimates $\mu(\bx)$ by $\hat{\mu}(\bx)=\sum_{d(\bX_i,\bx)\leq q_K}Y_i/K$ where $d(\cdot,\cdot)$ is a distance metric and $q_K$ is the $K$-th smallest values of $\{d(\bX_i,\bx),i=1,\cdots,n\}$. Ties can be broken randomly.
}
\medskip\noindent

\medskip\noindent
{\bf Support Vector Machines.} {\it The support vector machines (SVM) aims to separate the individuals from different classes with a hyperplane identified by a vector $(\omega_0,\boldsymbol{\omega}^\intercal)=(\omega_0,\omega_1,\cdots,\omega_p)$ so that the margin between the two classes can be maximized. Since the margin is proportional to $1/||\boldsymbol{\omega}||$, the naive SVM solves the optimization problem 
\begin{align*}
    \min_{\boldsymbol{\omega},\omega_0}||\boldsymbol{\omega}|| \quad s.t. \quad Y_i(\omega^\intercal\bX_i + \omega_0)\geq 1, i=1,\cdots,n.
\end{align*}
Samples on the boundary of the margin, i.e., those satisfy $\quad Y_i(\omega^\intercal\bX_i + \omega_0)= 1$, are called support vectors. When the two classes are not linearly separable, the SVM solves the following optimization problem
\begin{equation}
\label{eq:svm}
\begin{aligned}
    &\min_{\boldsymbol{\omega},\omega_0}||\boldsymbol{\omega}||^2/2 + C\sum_{i=1}^n\xi_i\\
    s.t. &\quad Y_i(\omega^\intercal\bX_i + \omega_0)\geq 1-\xi_i, \xi_i\geq 0, i=1,\cdots,n,
\end{aligned}
\end{equation}
where $\xi_1,\cdots,\xi_n$ are slack variables that allow the samples to be on the wrong side of the margin by the distance $\xi_i$, and $C$ is a cost parameter that controls the magnitude of $\sum_{i=1}^n\xi_i$. A dual problem of (\ref{eq:svm}) can be derived as
\begin{equation}
\label{eq:svm-dual}
\begin{aligned}
    &\max_{\alpha_1,\cdots,\alpha_n} \sum_{i=1}^n\alpha_i+ \sum_{i=1}^n\sum_{j=1}^n\alpha_i\alpha_jY_iY_j\bX_i^\intercal\bX_j\\
    s.t. &0\leq\alpha_i\leq C, i=1,\cdots,n, \sum_{i=1}^n\alpha_iY_i=0.
\end{aligned}
\end{equation}
A kernel trick is suggested by replacing $\bX_i^\intercal\bX_i$ with $K(\bX_i,\bX_j)$ where a Gaussian radial basis kernel corresponds to $K(\bX_i,\bX_j)=\exp(-\gamma||\bX_i-\bX_j||)$ so that the hyperplane will be built in a transformed space with high-dimensional features. The inverse distance of any observation with covariates $\bX_i$ to the decision hyperplane is proportional to $D_{i} = \sum_{j=1}^n\hat{\alpha}_jK(\bX_{i},\bX_j)$. The Platt scaling transformed these values into probabilities by doing logistic regression with the data $(D_i,Y_i),i=1,\cdots,n$.
}
\medskip\noindent

\medskip\noindent
{\bf Random Forests.} {\it A classification tree is a recursive binary tree that divides the covariates space into a bunch of partitions with its leaf nodes. The tree is usually constructed through two steps: (1) splitting the tree via splitting covariates by cutoff values based on certain rule such as Gini index or cross-entropy; the splitting will be stopped once the tree reaches maximum depth of the tree, maximum samples in each node, maximum number of nodes, or minimum information gains. (2) Pruning the tree based on errors of the tree on validation data and complexity of the tree. The estimated probability of a point will be the proportion of $Y=1$ in the node where the point is located.

Random forests for classification are an ensemble of classification trees. However, each tree in random forests is constructed based on a bootstrap sample and $\sqrt{p}$ of the original $p$ features. The estimated probability is the average of the estimated probabilities of the trees.
}
\medskip\noindent

\medskip\noindent
{\bf Gradient Boosting.} {\it The gradient boosting for classification assumes that $\mu(\bx)=logit^{-1}(f(\bx))$ for a function $f$ where $logit(x)=\log(x/(1-x))$. Here, we will assume that $Y_i\in\{0,1\}$. The model for $f$, denoted by $f_M(\bx)=\sum_{i=1}^M\eta h_m(\bx)$, is a weighted aggregation of a sequence of weak classifiers (e.g., trees). The algorithm learns the weak classifiers by sequentially minimizing the loss $L_n(f) = n^{-1}\sum_{i=1}^nY_i\log(\mu(\bX_i)) + (1-Y_i)\log(1-\mu(\bX_i)) = n^{-1}\sum_{i=1}^nY_if(\bX_i) - \log\{1+\exp(f(\bX_i))\}$. Let us begin by initializing the model $f_M$ with a constant prediction as
\[
f_0(x) = c = \log(\Bar{Y}/(1-\Bar{Y}))
\]
where $\Bar{Y}=n^{-1}\sum_{i=1}^nY_i$. At each iteration $m$, the model $f$ is updated by adding the new weak learner $h_m$:
\[
f_m(x) = f_{m-1}(x) + \eta \cdot h_m(x),
\]
where $\eta$ is the learning rate, a small constant that controls the contribution of each weak learner. For original gradient boosting, the weak learner $h_m$ is trained to predict the negative gradient of the loss function with respect to $f$ at $f_{m-1}$, i.e., $-\partial L_n(f_{m-1})/\partial f = \Bar{Y} - n^{-1}\sum_{i=1}^nlogit^{-1}(f_{m-1}(\bX_i))$. The XGBoost trains $h_m$ to predict the $-(\partial^2 L_n(f_{m-1})/\partial f\partial f^\intercal) \partial L_n(f_{m-1})/\partial f$, which matches the Newton's method. The algorithm continues iterating until a stopping criterion is met, such as (1) a predefined number of iterations is reached, or (2) the loss function no longer improves significantly. The final probability estimator will be $logit^{-1}(\sum_{i=1}^M\eta \hat{h}_m(\bx))$.
}
\medskip\noindent

\subsection*{A4. Additional Simulation Results}

\begin{figure}[t!]
    \centering
    \begin{subfigure}[b]{0.45\textwidth}
        \centering
        \includegraphics[width=\textwidth]{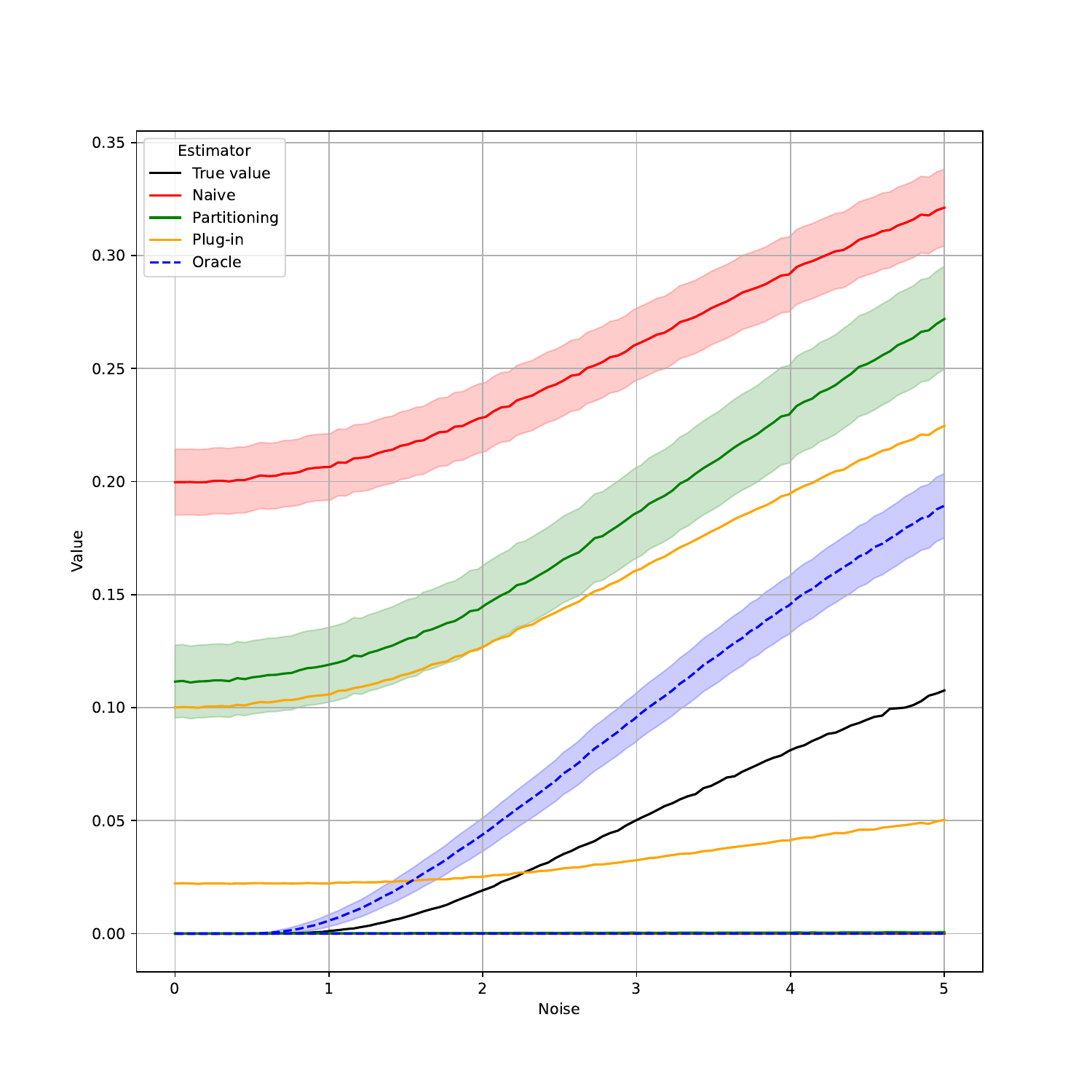} 
    \end{subfigure}
    \begin{subfigure}[b]{0.45\textwidth}
        \centering
        \includegraphics[width=\textwidth]{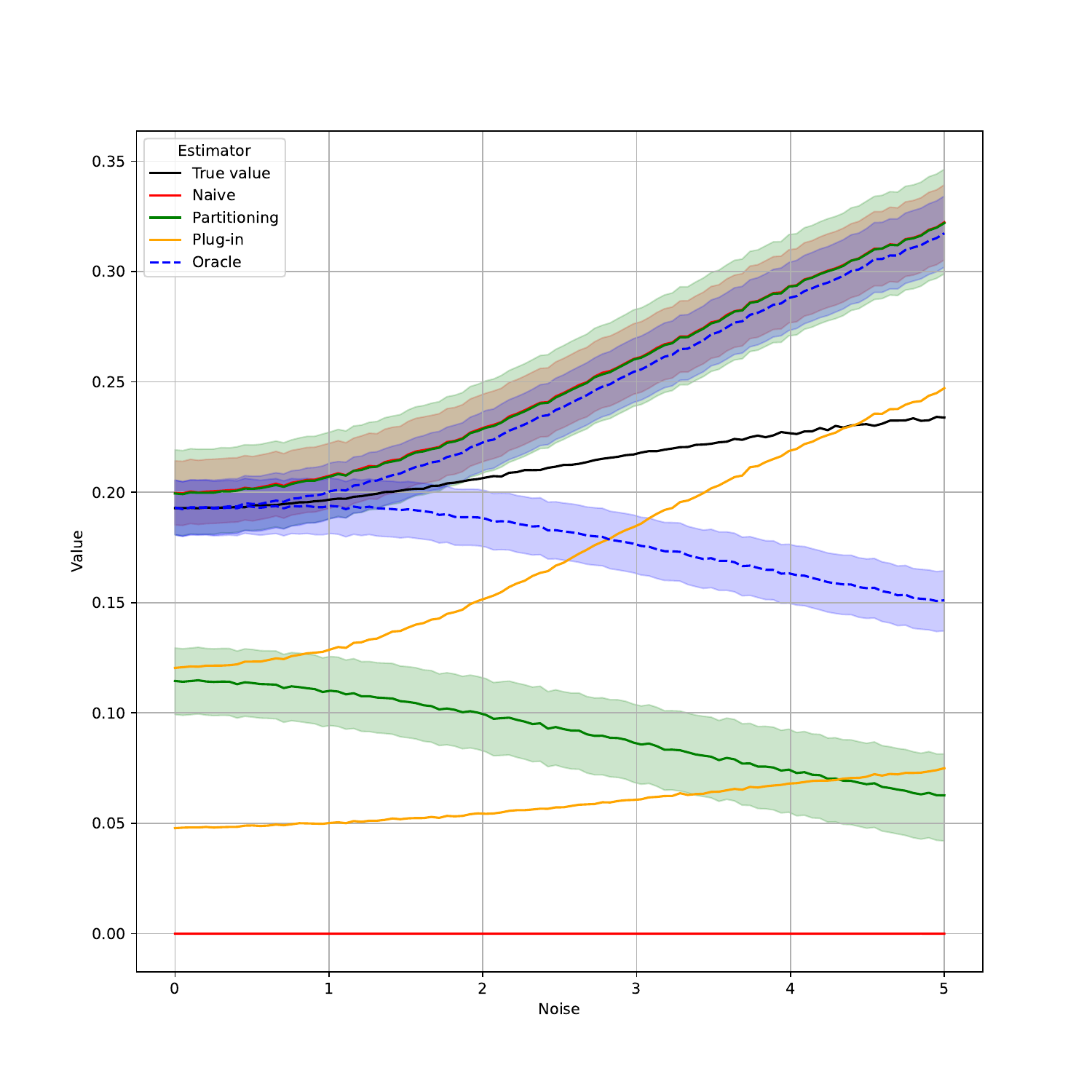} 
    \end{subfigure}
    \caption{Naive and Oracle bounds, estimated plug-in and partitioning-based bounds using the KNN algorithm, and their corresponding $75\%$ confidence bands. X-axis denotes the value of $\sigma_{\varepsilon}$. The plot for Scenario 1 is presented in the left panel, and the plot for Scenario 2 is presented in the right panel.}
    \label{fig:KNN}
\end{figure}

\begin{figure}[t!]
    \centering
    \begin{subfigure}[b]{0.45\textwidth}
        \centering
        \includegraphics[width=\textwidth]{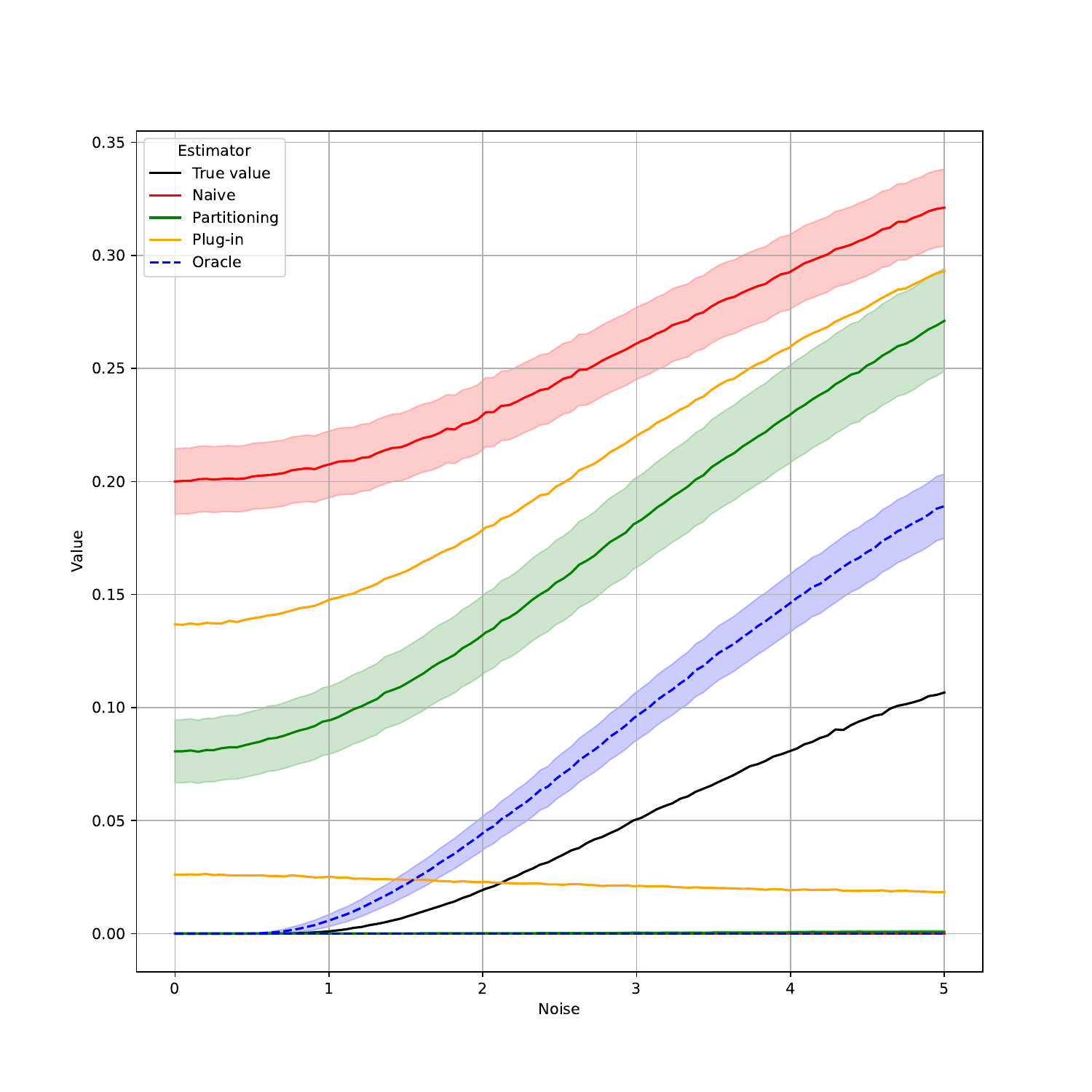} 
    \end{subfigure}
    \begin{subfigure}[b]{0.45\textwidth}
        \centering
        \includegraphics[width=\textwidth]{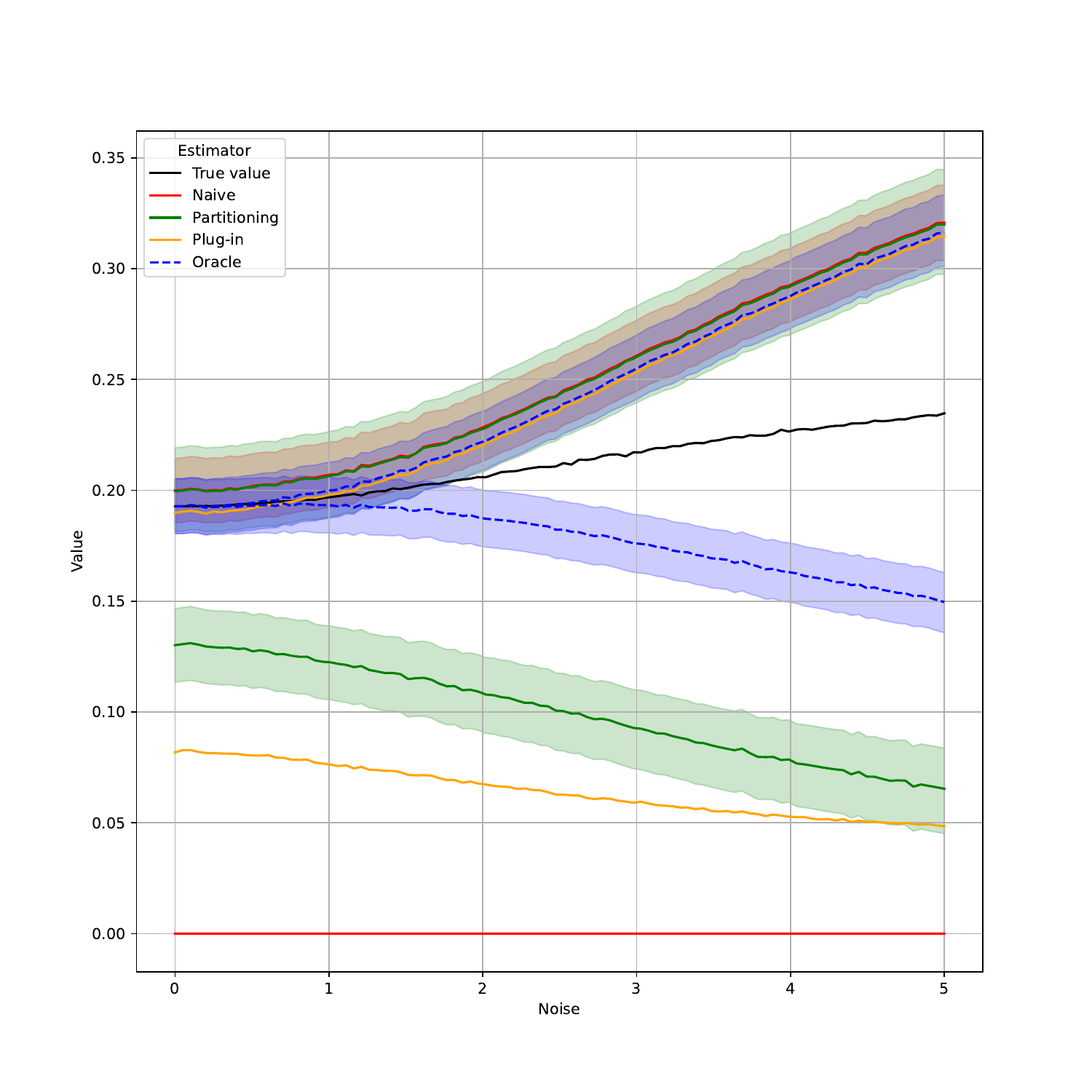} 
    \end{subfigure}
    \caption{Naive and Oracle bounds, estimated plug-in and partitioning-based bounds using the KNN algorithm combined with model calibration, and their corresponding $75\%$ confidence bands. X-axis denotes the value of $\sigma_{\varepsilon}$. The plot for Scenario 1 is presented in the left panel, and the plot for Scenario 2 is presented in the right panel.}
    \label{fig:KNN-calib}
\end{figure}

This section provides additional tables of the simulation results on (1) interval estimators of the six classification algorithms (Tables \ref{tab:estimation-setup1} and \ref{tab:estimation-setup2}); (2) confidence intervals and their coverage rates under the nominal levels $\alpha\in\{0.05,0.1,0.25,0.5\}$ for the Naive and Oracle bounds (Tables \ref{tab:inference-setup1} and \ref{tab:inference-setup2}); (3) 75\% extended confidence intervals and their coverage rates for $\theta$ (Table \ref{tab:extendedci}); (4) interval estimators of the six classification algorithms combined with model calibration (Tables \ref{tab:modelclibv-setup1} and \ref{tab:modelclibv-setup2}). This section also provides additional figures on the estimated lower and upper bounds under the KNN algorithm with $\sigma_{\varepsilon}$ varying from $0$ to $5$ (Figures \ref{fig:KNN} and \ref{fig:KNN-calib}). 

We briefly explain the results for the confidence intervals given in Tables \ref{tab:inference-setup1} and \ref{tab:inference-setup2}. Denote by $C_{\alpha,L}(\Delta)$ the confidence interval for the lower bound and $C_{\alpha,U}(\Delta)$ the one for the upper bound. We focus on the two partitions $\Delta_{0}$ and $\Delta_{full}$ because the true lower and upper bounds under them are known and computable. We consider the significance levels $\alpha\in\{0.05,0.1,0.25,0.5\}$ and the simulation results under the Scenario 1 and 2 are respectively summarized in Tables \ref{tab:inference-setup1} and \ref{tab:inference-setup2}. The LCovR and UCovR in the tables represent respectively the empirical coverage rate of $C_{\alpha,L}(\Delta)$ on $L(\theta,\Delta)$ and $C_{\alpha,U}(\Delta)$ on $U(\theta,\Delta)$. In general cases where $L(\theta,\Delta)$ or $U(\theta,\Delta)$ is away from zero, the empirical coverage rates of the proposed confidence intervals are close to the nominal levels. However, there are two cases to be noted and explained. First, under Scenario 1 with $\sigma_{\varepsilon}=1$ and $n=500$, the coverage rates of $C_{\alpha,U}(\Delta_{full})$ are away from the true nominal levels because there are too few samples in some cells of $\Delta_{full}$ to validate the confidence intervals constructed based on the asymptotic distributions of the estimated bounds in those cells. As $n$ increases to $2000$, the coverage rates are much better. Second, we can see that $C_{\alpha,L}(\Delta_0)$ over-covers $L(\theta,\Delta_0)$ with the coverage rate 100\% under all the setups. This is because the probability of $\widehat{L}(\theta,\Delta_0)$ is massed at zero under all the setups. 

\begin{table}[t!]
  \centering
  \caption{Monte Carlo simulation results for the interval estimation under \textbf{Scenario 1}. All values except the percentages in the table are presented in units scaled by $10^{-2}$.}
  \resizebox{0.8\textwidth}{!}{
    \begin{tabular}{lcclccrclccrcrr}
    \toprule
          &       &       & \multicolumn{4}{c}{Plug-in }   &       & \multicolumn{4}{c}{Partitioning} &       &       &  \\
\cmidrule{4-7}\cmidrule{9-12}    Method & $\theta$ &       & Estimate & Bias  & Width & CovR  &       & Estimate & Bias  & Width & CovR  &       & TMCR  & CMCR \\
    \midrule
    \multicolumn{15}{l}{$\sigma_{\varepsilon} = 1\quad n=500$} \\
    Naive & 0.1   &       & $[0.0, 20.5]$ & 0.0   & 20.5  & 100.0\% &       & $[0.0, 20.5]$ & 0.0   & 20.5  & 100.0\% &       & NaN   & NaN \\
    Logit & 0.1   &       & $[1.5, 17.9]$ & 1.4   & 16.3  & 0.4\% &       & $[0.3, 18.5]$ & 0.2   & 18.2  & 59.4\% &       & 35.1\% & 21.3\% \\
    NBayes & 0.1   &       & $[1.2, 12.2]$ & 1.1   & 11.0  & 0.3\% &       & $[0.1, 5.5]$ & 0.1   & 5.4   & 76.6\% &       & 18.7\% & 12.8\% \\
    SVM   & 0.1   &       & $[2.2, 15.9]$ & 2.1   & 13.7  & 0.0\% &       & $[0.2, 10.5]$ & 0.2   & 10.3  & 58.2\% &       & 26.7\% & 18.6\% \\
    KNN   & 0.1   &       & $[2.5, 11.7]$ & 2.4   & 9.2   & 0.0\% &       & $[0.1, 15.2]$ & 0.1   & 15.1  & 76.6\% &       & 31.3\% & 18.8\% \\
    RF    & 0.1   &       & $[0.8, 15.2]$ & 0.7   & 14.4  & 0.5\% &       & $[0.1, 5.3]$ & 0.1   & 5.2   & 76.5\% &       & 17.7\% & 10.6\% \\
    XGBoost & 0.1   &       & $[1.3, 4.1]$ & 1.2   & 2.8   & 0.0\% &       & $[0.1, 4.5]$ & 0.1   & 4.4   & 74.5\% &       & 17.5\% & 10.1\% \\
    Oracle & 0.1   &       & NaN   & NaN   & NaN   & NaN   &       & $[0.0, 0.6]$ & 0.0   & 0.6   & 73.5\% &       & NaN   & NaN \\
    \midrule
    \multicolumn{15}{l}{$\sigma_{\varepsilon} = 1\quad n=2000$} \\
    Naive & 0.1   &       & $[0.0, 20.7]$ & 0.0   & 20.7  & 100.0\% &        & $[0.0, 20.7]$ & 0.0   & 20.7  & 100.0\% &       & NaN   & NaN \\
    Logit & 0.1   &       & $[0.2, 18.9]$ & 0.1   & 18.8  & 43.2\% &       & $[0.0, 18.4]$ & 0.0   & 18.3  & 84.2\% &       & 33.0\% & 20.4\% \\
    NBayes & 0.1   &       & $[0.1, 12.8]$ & 0.0   & 12.6  & 47.1\% &       & $[0.0, 3.4]$ & 0.0   & 3.4   & 96.7\% &       & 13.9\% & 10.4\% \\
    SVM   & 0.1   &       & $[1.3, 7.5]$ & 1.2   & 6.2   & 0.0\% &       & $[0.0, 3.8]$ & 0.0   & 3.8   & 97.0\% &       & 17.4\% & 11.9\% \\
    KNN   & 0.1   &       & $[2.1, 10.7]$ & 2.0   & 8.6   & 0.0\% &       & $[0.0, 11.5]$ & 0.0   & 11.5  & 98.5\% &       & 26.1\% & 16.9\% \\
    RF    & 0.1   &       & $[0.4, 10.9]$ & 0.3   & 10.5  & 0.0\% &       & $[0.0, 3.2]$ & 0.0   & 3.2   & 94.9\% &       & 13.5\% & 7.6\% \\
    XGBoost & 0.1   &       & $[0.3, 1.3]$ & 0.2   & 1.0   & 1.1\% &       & $[0.0, 2.3]$ & 0.0   & 2.3   & 95.0\% &       & 13.1\% & 7.0\% \\
    Oracle & 0.1   &       & NaN   & NaN   & NaN   & NaN   &       & $[0.0, 0.6]$ & 0.0   & 0.6   & 98.1\% &       & NaN   & NaN \\
    \midrule
    \multicolumn{15}{l}{$\sigma_{\varepsilon} = 2\quad n=500$} \\
    Naive & 1.9   &       & $[0.0, 22.7]$ & 0.0   & 22.7  & 100.0\% &      & $[0.0, 22.7]$ & 0.0   & 22.7  & 100.0\% &       & NaN   & NaN \\
    Logit & 1.9   &       & $[1.5, 19.6]$ & 0.2   & 18.1  & 69.8\% &       & $[0.3, 20.7]$ & 0.0   & 20.4  & 99.5\% &       & 38.0\% & 23.7\% \\
    NBayes & 1.9   &       & $[1.4, 14.0]$ & 0.1   & 12.6  & 79.0\% &       & $[0.2, 8.8]$ & 0.0   & 8.6   & 99.6\% &       & 24.1\% & 15.1\% \\
    SVM   & 1.9   &       & $[2.0, 19.0]$ & 0.5   & 16.9  & 49.0\% &       & $[0.3, 14.0]$ & 0.0   & 13.7  & 98.7\% &       & 31.1\% & 20.8\% \\
    KNN   & 1.9   &       & $[2.7, 13.8]$ & 0.9   & 11.1  & 21.2\% &       & $[0.2, 17.7]$ & 0.0   & 17.5  & 99.2\% &       & 35.1\% & 21.1\% \\
    RF    & 1.9   &       & $[0.9, 17.7]$ & 0.0   & 16.8  & 95.8\% &       & $[0.2, 8.6]$ & 0.0   & 8.4   & 100.0\% &       & 23.8\% & 14.1\% \\
    XGBoost & 1.9   &       & $[2.2, 5.9]$ & 0.5   & 3.7   & 43.2\% &       & $[0.2, 8.5]$ & 0.0   & 8.3   & 99.7\% &       & 24.5\% & 14.2\% \\
    Oracle & 1.9   &       & NaN   & NaN   & NaN   & NaN   &       & $[0.0, 4.3]$ & 0.0   & 4.3   & 98.5\% &       & NaN   & NaN \\
    \midrule
    \multicolumn{15}{l}{$\sigma_{\varepsilon} = 2\quad n=2000$} \\
    Naive & 1.9   &       & $[0.0, 22.9]$ & 0.0   & 22.9  & 100.0\% &       & $[0.0, 22.9]$ & 0.0   & 22.9  & 100.0\% &       & NaN   & NaN \\
    Logit & 1.9   &       & $[0.1, 20.7]$ & 0.0   & 20.6  & 100.0\% &       & $[0.0, 20.6]$ & 0.0   & 20.6  & 100.0\% &       & 35.7\% & 22.6\% \\
    NBayes & 1.9   &       & $[0.1, 14.6]$ & 0.0   & 14.4  & 100.0\% &       & $[0.0, 6.8]$ & 0.0   & 6.8   & 100.0\% &       & 19.8\% & 12.6\% \\
    SVM   & 1.9   &       & $[1.5, 11.4]$ & 0.0   & 9.9   & 86.6\% &       & $[0.0, 7.6]$ & 0.0   & 7.6   & 100.0\% &       & 23.1\% & 14.6\% \\
    KNN   & 1.9   &       & $[2.4, 12.9]$ & 0.5   & 10.5  & 9.9\% &       & $[0.0, 14.3]$ & 0.0   & 14.3  & 100.0\% &       & 30.6\% & 19.2\% \\
    RF    & 1.9   &       & $[0.4, 13.6]$ & 0.0   & 13.2  & 100.0\% &       & $[0.0, 6.6]$ & 0.0   & 6.6   & 100.0\% &       & 19.7\% & 11.0\% \\
    XGBoost & 1.9   &       & $[0.8, 2.7]$ & 0.0   & 1.9   & 98.3\% &       & $[0.0, 6.3]$ & 0.0   & 6.3   & 100.0\% &       & 20.6\% & 11.2\% \\
    Oracle & 1.9   &       & NaN   & NaN   & NaN   & NaN   &       & $[0.0, 4.4]$ & 0.0   & 4.4   & 100.0\% &       & NaN   & NaN \\
    \bottomrule
    \end{tabular}%
    }
  \label{tab:estimation-setup1}%
\end{table}%

\begin{table}[t!]
  \centering
  \caption{Monte Carlo simulation results for the interval estimation under \textbf{Scenario 2}. All values except the percentages in the table are presented in units scaled by $10^{-2}$.}
  \resizebox{0.8\textwidth}{!}{
    \begin{tabular}{lcclccrclccrcrr}
    \toprule
          &       &       & \multicolumn{4}{c}{Plug-in}   &       & \multicolumn{4}{c}{Partitioning} &       &       &  \\
\cmidrule{4-7}\cmidrule{9-12}    Method & $\theta$ &       & Estimate & Bias  & Width & CovR  &       & Estimate & Bias  & Width & CovR  &       & TMCR  & CMCR \\
    \midrule
    \multicolumn{15}{l}{$\sigma_{\varepsilon} = 1\quad n=500$} \\
    Naive & 19.7  &       & $[0.0, 20.6]$ & 0.6   & 20.6  & 61.6\% &       & $[0.0, 20.6]$ & 0.6   & 20.6  & 61.6\% &       & NaN   & NaN \\
    Logit & 19.7  &       & $[3.7, 19.9]$ & 0.9   & 16.2  & 51.7\% &       & $[1.9, 20.3]$ & 0.7   & 18.4  & 58.4\% &       & 39.3\% & 21.4\% \\
    NBayes & 19.7  &       & $[13.2, 21.3]$ & 0.5   & 8.0   & 70.6\% &       & $[17.0, 20.5]$ & 0.7   & 3.5   & 50.7\% &       & 23.1\% & 12.8\% \\
    SVM   & 19.7  &       & $[6.2, 20.4]$ & 0.6   & 14.1  & 60.7\% &       & $[10.5, 20.4]$ & 0.6   & 9.9   & 61.4\% &       & 32.4\% & 18.6\% \\
    KNN   & 19.7  &       & $[3.0, 12.4]$ & 7.2   & 9.5   & 0.1\% &       & $[7.3, 20.4]$ & 0.6   & 13.2  & 62.3\% &       & 35.9\% & 18.7\% \\
    RF    & 19.7  &       & $[7.6, 20.5]$ & 0.5   & 12.9  & 64.2\% &       & $[16.7, 20.4]$ & 0.7   & 3.7   & 49.5\% &       & 21.7\% & 10.6\% \\
    XGBoost & 19.7  &       & $[16.2, 17.9]$ & 2.2   & 1.7   & 13.6\% &       & $[17.6, 20.4]$ & 0.8   & 2.9   & 43.8\% &       & 20.0\% & 10.0\% \\
    Oracle & 19.7  &       & NaN   & NaN   & NaN   & NaN   &       & $[19.2, 19.8]$ & 1.4   & 0.6   & 11.4\% &       & NaN   & NaN \\
    \midrule
    \multicolumn{15}{l}{$\sigma_{\varepsilon} = 1\quad n=2000$} \\
    Naive & 19.7  &       & $[0.0, 20.7]$ & 0.2   & 20.7  & 78.1\% &       & $[0.0, 20.7]$ & 0.2   & 20.7  & 78.1\% &       & NaN   & NaN \\
    Logit & 19.7  &       & $[2.2, 20.5]$ & 0.2   & 18.3  & 73.9\% &       & $[1.7, 20.6]$ & 0.2   & 18.8  & 75.9\% &       & 37.1\% & 20.4\% \\
    NBayes & 19.7  &       & $[13.3, 22.3]$ & 0.0   & 8.9   & 96.5\% &       & $[18.7, 20.6]$ & 0.2   & 1.9   & 62.4\% &       & 19.1\% & 10.4\% \\
    SVM   & 19.7  &       & $[13.5, 19.9]$ & 0.3   & 6.4   & 58.1\% &       & $[16.3, 20.5]$ & 0.1   & 4.2   & 78.7\% &       & 24.5\% & 11.9\% \\
    KNN   & 19.7  &       & $[4.8, 12.8]$ & 6.9   & 8.1   & 0.0\% &       & $[11.3, 20.6]$ & 0.2   & 9.3   & 79.2\% &       & 30.6\% & 16.9\% \\
    RF    & 19.7  &       & $[11.7, 20.5]$ & 0.1   & 8.8   & 80.0\% &       & $[18.4, 20.6]$ & 0.2   & 2.1   & 66.9\% &       & 16.5\% & 7.6\% \\
    XGBoost & 19.7  &       & $[18.7, 19.2]$ & 0.8   & 0.5   & 14.9\% &       & $[18.7, 20.5]$ & 0.3   & 1.8   & 58.1\% &       & 16.3\% & 7.1\% \\
    Oracle & 19.7  &       & NaN   & NaN   & NaN   & NaN   &       & $[19.3, 19.9]$ & 0.6   & 0.7   & 24.8\% &       & NaN   & NaN \\
    \midrule
    \multicolumn{15}{l}{$\sigma_{\varepsilon} = 2\quad n=500$} \\
    Naive & 20.6  &       & $[0.0, 22.6]$ & 0.3   & 22.6  & 75.9\% &       & $[0.0, 22.6]$ & 0.3   & 22.6  & 75.9\% &       & NaN   & NaN \\
    Logit & 20.6  &       & $[4.1, 21.8]$ & 0.5   & 17.6  & 65.5\% &       & $[2.0, 22.2]$ & 0.4   & 20.2  & 70.5\% &       & 40.2\% & 23.6\% \\
    NBayes & 20.6  &       & $[13.2, 23.0]$ & 0.3   & 9.8   & 79.1\% &       & $[15.9, 22.4]$ & 0.3   & 6.5   & 75.5\% &       & 28.4\% & 15.2\% \\
    SVM   & 20.6  &       & $[5.2, 22.2]$ & 0.4   & 17.0  & 74.5\% &       & $[9.1, 22.3]$ & 0.4   & 13.2  & 74.5\% &       & 35.2\% & 20.6\% \\
    KNN   & 20.6  &       & $[3.5, 14.6]$ & 6.1   & 11.1  & 1.3\% &       & $[6.2, 22.4]$ & 0.3   & 16.2  & 74.7\% &       & 38.5\% & 21.0\% \\
    RF    & 20.6  &       & $[7.4, 22.5]$ & 0.3   & 15.1  & 78.5\% &       & $[15.6, 22.3]$ & 0.3   & 6.7   & 73.7\% &       & 28.6\% & 14.0\% \\
    XGBoost & 20.6  &       & $[16.2, 18.8]$ & 2.3   & 2.6   & 19.6\% &       & $[16.0, 22.4]$ & 0.3   & 6.4   & 73.5\% &       & 27.7\% & 14.3\% \\
    Oracle & 20.6  &       & NaN   & NaN   & NaN   & NaN   &       & $[18.6, 21.8]$ & 0.6   & 3.3   & 51.9\% &       & NaN   & NaN \\
    \midrule
    \multicolumn{15}{l}{$\sigma_{\varepsilon} = 2\quad n=2000$} \\
    Naive & 20.6  &       & $[0.0, 23.0]$ & 0.0   & 23.0  & 95.4\% &       & $[0.0, 23.0]$ & 0.0   & 23.0  & 95.4\% &       & NaN   & NaN \\
    Logit & 20.6  &       & $[2.5, 22.8]$ & 0.0   & 20.2  & 94.9\% &       & $[1.8, 22.8]$ & 0.0   & 21.0  & 94.5\% &       & 37.7\% & 22.6\% \\
    NBayes & 20.6  &       & $[13.2, 24.3]$ & 0.0   & 11.1  & 99.5\% &       & $[18.1, 22.8]$ & 0.0   & 4.7   & 95.4\% &       & 25.2\% & 12.6\% \\
    SVM   & 20.6  &       & $[11.9, 22.0]$ & 0.1   & 10.1  & 86.8\% &       & $[15.3, 22.8]$ & 0.0   & 7.5   & 96.5\% &       & 29.6\% & 14.6\% \\
    KNN   & 20.6  &       & $[5.2, 15.2]$ & 5.5   & 10.0  & 0.0\% &       & $[10.0, 22.9]$ & 0.0   & 12.9  & 95.8\% &       & 34.8\% & 19.2\% \\
    RF    & 20.6  &       & $[11.5, 22.8]$ & 0.0   & 11.3  & 96.1\% &       & $[17.9, 22.8]$ & 0.0   & 4.9   & 95.8\% &       & 24.2\% & 11.1\% \\
    XGBoost & 20.6  &       & $[19.2, 20.4]$ & 0.7   & 1.2   & 28.3\% &       & $[17.7, 22.8]$ & 0.0   & 5.1   & 95.9\% &       & 24.8\% & 11.3\% \\
    Oracle & 20.6  &       & NaN   & NaN   & NaN   & NaN   &       & $[18.8, 22.3]$ & 0.1   & 3.5   & 88.0\% &       & NaN   & NaN \\
    \bottomrule
    \end{tabular}%
    }
  \label{tab:estimation-setup2}%
\end{table}%

\begin{table}[t!]
  \centering
  \caption{Monte Carlo simulation results for the confidence intervals under \textbf{Scenario 1}. All values except the percentages in the table are presented in units scaled by $10^{-2}$.}
  \resizebox{\textwidth}{!}{
    \begin{tabular}{lclrclrrclrclr}
    \toprule
          & \multicolumn{6}{c}{Naive}                     &       & \multicolumn{6}{c}{Oracle} \\
\cmidrule{2-7}\cmidrule{9-14}    $\alpha$ & $L(\theta,\Delta_{0})$ & $C_{\alpha,L}(\Delta_0)$ & LCovR & $U(\theta,\Delta_{0})$ & $C_{\alpha,U}(\Delta_0)$ & UCovR &       & $L(\theta,\Delta_{full})$ & $C_{\alpha,L}(\Delta_{full})$ & LCovR & $U(\theta,\Delta_{full})$ & $C_{\alpha,U}(\Delta_{full})$ & UCovR  \\
    \midrule
    \multicolumn{14}{l}{$\sigma_{\varepsilon} = 1\quad n=500$} \\
    0.05  & 0.0    & $[0.0, 0.0]$ & 100.0\% & 20.7  & $[15.7, 25.8]$ & 94.1\% &       & 0.0    & $[0.0, 0.0]$ & 100.0\% & 0.6   & $[0.0, 1.4]$ & 74.0\% \\
    0.10  & 0.0    & $[0.0, 0.0]$ & 100.0\% & 20.7  & $[16.5, 25.0]$ & 89.8\% &       & 0.0    & $[0.0, 0.0]$ & 100.0\% & 0.6   & $[0.0, 1.2]$ & 73.6\% \\
    0.25  & 0.0    & $[0.0, 0.0]$ & 100.0\% & 20.7  & $[17.8, 23.7]$ & 76.1\% &       & 0.0    & $[0.0, 0.0]$ & 100.0\% & 0.6   & $[0.1, 1.0]$ & 67.3\% \\
    0.50  & 0.0    & $[0.0, 0.0]$ & 100.0\% & 20.7  & $[19.0, 22.5]$ & 51.3\% &       & 0.0    & $[0.0, 0.0]$ & 100.0\% & 0.6   & $[0.3, 0.8]$ & 55.3\% \\
    \midrule
    \multicolumn{14}{l}{$\sigma_{\varepsilon} = 1\quad n=2000$} \\
    0.05  & 0.0    & $[0.0, 0.0]$ & 100.0\% & 20.7  & $[18.2, 23.2]$ & 95.1\% &       & 0.0    & $[0.0, 0.0]$ & 100.0\% & 0.6   & $[0.1, 1.0]$ & 90.7\% \\
    0.10  & 0.0    & $[0.0, 0.0]$ & 100.0\% & 20.7  & $[18.6, 22.8]$ & 90.8\% &       & 0.0    & $[0.0, 0.0]$ & 100.0\% & 0.6   & $[0.2, 0.9]$ & 88.5\% \\
    0.25  & 0.0    & $[0.0, 0.0]$ & 100.0\% & 20.7  & $[19.2, 22.2]$ & 75.2\% &       & 0.0    & $[0.0, 0.0]$ & 100.0\% & 0.6   & $[0.3, 0.8]$ & 73.4\% \\
    0.50  & 0.0    & $[0.0, 0.0]$ & 100.0\% & 20.7  & $[19.8, 21.5]$ & 50.3\% &       & 0.0    & $[0.0, 0.0]$ & 100.0\% & 0.6   & $[0.4, 0.7]$ & 43.6\% \\
    \midrule
    \multicolumn{14}{l}{$\sigma_{\varepsilon} = 2\quad n=500$} \\
    0.05  & 0.0    & $[0.0, 0.0]$ & 100.0\% & 22.8  & $[17.5, 27.9]$ & 94.3\% &       & 0.0    & $[0.0, 0.0]$ & 100.0\% & 4.3   & $[1.9, 6.9]$ & 92.5\% \\
    0.10  & 0.0    & $[0.0, 0.0]$ & 100.0\% & 22.8  & $[18.4, 27.1]$ & 91.6\% &       & 0.0    & $[0.0, 0.0]$ & 100.0\% & 4.3   & $[2.3, 6.5]$ & 89.5\% \\
    0.25  & 0.0    & $[0.0, 0.0]$ & 100.0\% & 22.8  & $[19.7, 25.8]$ & 76.0\% &       & 0.0    & $[0.0, 0.0]$ & 100.0\% & 4.3   & $[2.9, 5.9]$ & 74.8\% \\
    0.50  & 0.0    & $[0.0, 0.0]$ & 100.0\% & 22.8  & $[20.9, 24.5]$ & 51.3\% &       & 0.0    & $[0.0, 0.0]$ & 100.0\% & 4.3   & $[3.5, 5.2]$ & 48.4\% \\
    \midrule
    \multicolumn{14}{l}{$\sigma_{\varepsilon} = 2\quad n=2000$} \\
    0.05  & 0.0    & $[0.0, 0.0]$ & 100.0\% & 23.0  & $[20.3, 25.5]$ & 95.2\% &       & 0.0    & $[0.0, 0.0]$ & 100.0\% & 4.5   & $[3.2, 5.7]$ & 94.1\% \\
    0.10  & 0.0    & $[0.0, 0.0]$ & 100.0\% & 23.0  & $[20.7, 25.0]$ & 90.2\% &       & 0.0    & $[0.0, 0.0]$ & 100.0\% & 4.5   & $[3.4, 5.5]$ & 90.7\% \\
    0.25  & 0.0    & $[0.0, 0.0]$ & 100.0\% & 23.0  & $[21.3, 24.4]$ & 77.8\% &       & 0.0    & $[0.0, 0.0]$ & 100.0\% & 4.5   & $[3.7, 5.2]$ & 76.9\% \\
    0.50  & 0.0    & $[0.0, 0.0]$ & 100.0\% & 23.0  & $[22.0, 23.8]$ & 53.7\% &       & 0.0    & $[0.0, 0.0]$ & 100.0\% & 4.5   & $[4.0, 4.9]$ & 50.0\% \\
    \bottomrule
    \end{tabular}%
    }
  \label{tab:inference-setup1}%
\end{table}%

\begin{table}[t!]
  \centering
  \caption{Monte Carlo simulation results for the confidence intervals under \textbf{Scenario 2}. All values except the percentages in the table are presented in units scaled by $10^{-2}$.}
  \resizebox{\textwidth}{!}{
    \begin{tabular}{lclrclrrclrclr}
    \toprule
          & \multicolumn{6}{c}{Naive}                     &       & \multicolumn{6}{c}{Oracle} \\
\cmidrule{2-7}\cmidrule{9-14}    $\alpha$ & $L(\theta,\Delta_{0})$ & $C_{\alpha,L}(\Delta_0)$ & LCovR & $U(\theta,\Delta_{0})$ & $C_{\alpha,U}(\Delta_0)$ & UCovR &       & $L(\theta,\Delta_{full})$ & $C_{\alpha,L}(\Delta_{full})$ & LCovR & $U(\theta,\Delta_{full})$ & $C_{\alpha,U}(\Delta_{full})$ & UCovR  \\
    \midrule
    \multicolumn{14}{l}{$\sigma_{\varepsilon} = 1\quad n=500$} \\
    0.05  & 0.0    & $[0.0, 0.0]$ & 100.0\% & 20.7  & $[15.5, 25.5]$ & 94.4\% &       & 19.3  & $[14.2, 23.3]$ & 94.1\% & 20.0    & $[15.4, 24.6]$ & 95.2\% \\
    0.10  & 0.0    & $[0.0, 0.0]$ & 100.0\% & 20.7  & $[16.3, 24.7]$ & 89.6\% &       & 19.3  & $[15.2, 22.6]$ & 89.9\% & 20.0    & $[16.1, 23.6]$ & 89.5\% \\
    0.25  & 0.0    & $[0.0, 0.0]$ & 100.0\% & 20.7  & $[17.6, 23.4]$ & 78.0\% &       & 19.3  & $[16.4, 21.5]$ & 73.5\% & 20.0    & $[17.2, 22.4]$ & 77.1\% \\
    0.50  & 0.0    & $[0.0, 0.0]$ & 100.0\% & 20.7  & $[18.8, 22.2]$ & 52.6\% &       & 19.3  & $[17.5, 20.5]$ & 49.7\% & 20.0    & $[18.3, 21.3]$ & 52.9\% \\
    \midrule
    \multicolumn{14}{l}{$\sigma_{\varepsilon} = 1\quad n=2000$} \\
    0.05  & 0.0    & $[0.0, 0.0]$ & 100.0\% & 20.7  & $[18.2, 23.3]$ & 94.7\% &       & 19.3  & $[17.2, 21.5]$ & 94.3\% & 20.0    & $[17.9, 22.2]$ & 94.7\% \\
    0.10  & 0.0    & $[0.0, 0.0]$ & 100.0\% & 20.7  & $[18.6, 22.9]$ & 89.0\% &       & 19.3  & $[17.6, 21.2]$ & 89.8\% & 20.0    & $[18.3, 21.9]$ & 88.9\% \\
    0.25  & 0.0    & $[0.0, 0.0]$ & 100.0\% & 20.7  & $[19.3, 22.2]$ & 74.8\% &       & 19.3  & $[18.1, 20.6]$ & 76.1\% & 20.0    & $[18.8, 21.3]$ & 75.1\% \\
    0.50  & 0.0    & $[0.0, 0.0]$ & 100.0\% & 20.7  & $[19.9, 21.6]$ & 50.6\% &       & 19.3  & $[18.6, 20.1]$ & 49.4\% & 20.0    & $[19.3, 20.8]$ & 49.7\% \\
    \midrule
    \multicolumn{14}{l}{$\sigma_{\varepsilon} = 2\quad n=500$} \\
    0.05  & 0.0    & $[0.0, 0.0]$ & 100.0\% & 22.8  & $[17.8, 28.2]$ & 95.4\% &       & 18.7  & $[13.6, 23.1]$ & 96.8\% & 22.2  & $[17.7, 27.7]$ & 96.4\% \\
    0.10  & 0.0    & $[0.0, 0.0]$ & 100.0\% & 22.8  & $[18.6, 27.4]$ & 90.7\% &       & 18.7  & $[14.7, 22.4]$ & 92.1\% & 22.2  & $[18.4, 26.5]$ & 90.7\% \\
    0.25  & 0.0    & $[0.0, 0.0]$ & 100.0\% & 22.8  & $[19.9, 26.1]$ & 75.0\% &       & 18.7  & $[16.1, 21.3]$ & 77.2\% & 22.2  & $[19.6, 25.1]$ & 75.3\% \\
    0.50  & 0.0    & $[0.0, 0.0]$ & 100.0\% & 22.8  & $[21.2, 24.8]$ & 47.6\% &       & 18.7  & $[17.2, 20.3]$ & 50.0\% & 22.2  & $[20.7, 23.9]$ & 50.0\% \\
    \midrule
    \multicolumn{14}{l}{$\sigma_{\varepsilon} = 2\quad n=2000$} \\
    0.05  & 0.0    & $[0.0, 0.0]$ & 100.0\% & 22.8  & $[20.3, 25.6]$ & 94.8\% &       & 18.8  & $[16.6, 21.0]$ & 95.3\% & 22.2  & $[20.0, 24.6]$ & 94.9\% \\
    0.10  & 0.0    & $[0.0, 0.0]$ & 100.0\% & 22.8  & $[20.8, 25.1]$ & 90.0\% &       & 18.8  & $[17.0, 20.7]$ & 90.1\% & 22.2  & $[20.4, 24.2]$ & 90.7\% \\
    0.25  & 0.0    & $[0.0, 0.0]$ & 100.0\% & 22.8  & $[21.4, 24.5]$ & 75.0\% &       & 18.8  & $[17.5, 20.1]$ & 74.2\% & 22.2  & $[21.0, 23.7]$ & 75.1\% \\
    0.50  & 0.0    & $[0.0, 0.0]$ & 100.0\% & 22.8  & $[22.1, 23.8]$ & 49.4\% &       & 18.8  & $[18.1, 19.6]$ & 50.4\% & 22.2  & $[21.5, 23.1]$ & 49.7\% \\
    \bottomrule
    \end{tabular}%
    }
  \label{tab:inference-setup2}%
\end{table}%

\begin{table}[t!]
  \centering
  \caption{Monte Carlo simulation results on the extended confidence intervals with $\alpha=0.25$. All values except the percentages in the table are presented in units scaled by $10^{-2}$.}
  \resizebox{0.7\textwidth}{!}{
    \begin{tabular}{lcclccrrlccr}
    \toprule
          &       &       & \multicolumn{4}{c}{Scenario 1}    &       & \multicolumn{4}{c}{Scenario 2} \\
\cmidrule{4-7}\cmidrule{9-12}    Method & $\theta$ &       & Estimate & Bias  & Width & CovR  &       & Estimate & Bias  & Width & CovR \\
    \midrule
    \multicolumn{12}{l}{$\sigma_{\varepsilon} = 1\quad n=500$} \\
    Naive & 19.7  &       & [0.0, 23.5] & 0.0   & 23.5  & 100.00\% &       & [0.0, 23.6] & 0.1   & 23.6  & 93.20\% \\
    Logit & 19.7  &       & [0.0, 23.0] & 0.0   & 23.0  & 95.40\% &       & [0.3, 25.0] & 0.0   & 24.6  & 97.30\% \\
    NBayes & 19.7  &       & [0.0, 7.9] & 0.0   & 7.9   & 98.60\% &       & [13.0, 24.5] & 0.0   & 11.5  & 98.30\% \\
    SVM   & 19.7  &       & [0.0, 13.8] & 0.0   & 13.8  & 97.40\% &       & [7.0, 24.6] & 0.0   & 17.7  & 98.40\% \\
    KNN   & 19.7  &       & [0.0, 19.0] & 0.0   & 19.0  & 98.70\% &       & [4.0, 24.6] & 0.0   & 20.7  & 97.20\% \\
    RF    & 19.7  &       & [0.0, 7.7] & 0.0   & 7.7   & 97.60\% &       & [13.2, 24.2] & 0.0   & 11.0  & 96.80\% \\
    XGBoost & 19.7  &       & [0.0, 6.6] & 0.0   & 6.6   & 98.40\% &       & [13.8, 24.4] & 0.0   & 10.5  & 97.00\% \\
    Oracle & 19.7  &       & [0.0, 1.0] & 0.0   & 1.0   & 73.50\% &       & [16.6, 22.6] & 0.2   & 6.0   & 82.80\% \\
    \midrule
    \multicolumn{12}{l}{$\sigma_{\varepsilon} = 1\quad n=2000$} \\
    Naive & 19.7  &       & [0.0, 22.2] & 0.0   & 22.2  & 100.00\% &       & [0.0, 22.2] & 0.0   & 22.2  & 96.90\% \\
    Logit & 19.7  &       & [0.0, 20.4] & 0.0   & 20.4  & 98.70\% &       & [0.6, 22.8] & 0.0   & 22.2  & 98.90\% \\
    NBayes & 19.7  &       & [0.0, 4.3] & 0.0   & 4.3   & 99.90\% &       & [16.8, 22.5] & 0.0   & 5.7   & 98.90\% \\
    SVM   & 19.7  &       & [0.0, 4.7] & 0.0   & 4.7   & 100.00\% &       & [14.6, 22.4] & 0.0   & 7.9   & 99.00\% \\
    KNN   & 19.7  &       & [0.0, 13.2] & 0.0   & 13.2  & 99.90\% &       & [9.6, 22.6] & 0.0   & 13.0  & 99.00\% \\
    RF    & 19.7  &       & [0.0, 4.1] & 0.0   & 4.1   & 99.80\% &       & [16.7, 22.4] & 0.0   & 5.7   & 98.90\% \\
    XGBoost & 19.7  &       & [0.0, 3.1] & 0.0   & 3.1   & 100.00\% &       & [16.9, 22.4] & 0.0   & 5.4   & 98.80\% \\
    Oracle & 19.7  &       & [0.0, 0.8] & 0.0   & 0.8   & 99.60\% &       & [18.0, 21.2] & 0.1   & 3.2   & 85.10\% \\
    \midrule
    \multicolumn{12}{l}{$\sigma_{\varepsilon} = 2\quad n=500$} \\
    Naive & 20.6  &       & [0.0, 25.8] & 0.0   & 25.8  & 100.00\% &       & [0.0, 25.6] & 0.0   & 25.6  & 97.60\% \\
    Logit & 20.6  &       & [0.0, 25.5] & 0.0   & 25.5  & 100.00\% &       & [0.3, 27.0] & 0.0   & 26.7  & 99.40\% \\
    NBayes & 20.6  &       & [0.0, 11.9] & 0.0   & 11.9  & 100.00\% &       & [11.7, 26.8] & 0.0   & 15.0  & 99.40\% \\
    SVM   & 20.6  &       & [0.0, 17.9] & 0.0   & 17.8  & 100.00\% &       & [5.7, 26.7] & 0.0   & 21.0  & 99.50\% \\
    KNN   & 20.6  &       & [0.0, 21.8] & 0.0   & 21.8  & 100.00\% &       & [2.8, 26.8] & 0.0   & 24.0  & 99.30\% \\
    RF    & 20.6  &       & [0.0, 11.6] & 0.0   & 11.6  & 100.00\% &       & [11.9, 26.5] & 0.0   & 14.6  & 99.20\% \\
    XGBoost & 20.6  &       & [0.0, 11.4] & 0.0   & 11.4  & 100.00\% &       & [11.9, 26.7] & 0.0   & 14.8  & 99.50\% \\
    Oracle & 20.6  &       & [0.0, 5.8] & 0.0   & 5.8   & 99.80\% &       & [15.8, 24.8] & 0.0   & 9.0   & 95.10\% \\
    \midrule
    \multicolumn{12}{l}{$\sigma_{\varepsilon} = 2\quad n=2000$} \\
    Naive & 20.6  &       & [0.0, 24.4] & 0.0   & 24.4  & 100.00\% &       & [0.0, 24.5] & 0.0   & 24.5  & 99.70\% \\
    Logit & 20.6  &       & [0.0, 22.8] & 0.0   & 22.8  & 100.00\% &       & [0.6, 25.1] & 0.0   & 24.5  & 99.80\% \\
    NBayes & 20.6  &       & [0.0, 8.1] & 0.0   & 8.1   & 100.00\% &       & [16.1, 24.9] & 0.0   & 8.7   & 100.00\% \\
    SVM   & 20.6  &       & [0.0, 9.0] & 0.0   & 9.0   & 100.00\% &       & [13.5, 24.9] & 0.0   & 11.3  & 100.00\% \\
    KNN   & 20.6  &       & [0.0, 16.1] & 0.0   & 16.1  & 100.00\% &       & [8.2, 25.0] & 0.0   & 16.8  & 99.60\% \\
    RF    & 20.6  &       & [0.0, 7.9] & 0.0   & 7.9   & 100.00\% &       & [16.1, 24.8] & 0.0   & 8.8   & 100.00\% \\
    XGBoost & 20.6  &       & [0.0, 7.6] & 0.0   & 7.6   & 100.00\% &       & [15.9, 24.8] & 0.0   & 9.0   & 100.00\% \\
    Oracle & 20.6  &       & [0.0, 5.2] & 0.0   & 5.2   & 100.00\% &       & [17.6, 23.7] & 0.0   & 6.1   & 99.00\% \\
    \bottomrule
    \end{tabular}%
    }
  \label{tab:extendedci}%
\end{table}%

\begin{table}[t!]
  \centering
  \caption{Monte Carlo simulation results on the interval estimation combined with model calibration under \textbf{Scenario 1}. All values except the percentages in the table are presented in units scaled by $10^{-2}$.}
  \resizebox{\textwidth}{!}{
    \begin{tabular}{lcrlccrrlccrrrr}
    \toprule
          &       &       & \multicolumn{4}{c}{Plug-in}   &       & \multicolumn{4}{c}{Partitioning} &       &       &  \\
\cmidrule{4-7}\cmidrule{9-12}    Method & $\theta$ &       & Estimate & Bias  & Width & CovR  &       & Estimate & Bias  & Width & CovR  &       & TMCR  & CMCR \\
    \midrule
    \multicolumn{15}{l}{$\sigma_{\varepsilon} = 1\quad n=500$} \\
    Logit & 0.1   &       & [1.6, 18.0] & 1.5   & 16.4  & 0.10\% &       & [0.3, 18.4] & 0.2   & 18.2  & 57.60\% &       & 35.10\% & 21.30\% \\
    NBayes & 0.1   &       & [2.3, 9.2] & 2.2   & 6.9   & 0.00\% &       & [0.1, 5.4] & 0.1   & 5.3   & 76.70\% &       & 18.70\% & 12.80\% \\
    SVM   & 0.1   &       & [3.0, 13.4] & 2.9   & 10.4  & 0.00\% &       & [0.2, 10.4] & 0.2   & 10.2  & 60.60\% &       & 26.70\% & 18.60\% \\
    KNN   & 0.1   &       & [2.0, 18.1] & 1.9   & 16.0  & 0.00\% &       & [0.2, 14.7] & 0.2   & 14.5  & 57.70\% &       & 31.30\% & 18.80\% \\
    RF    & 0.1   &       & [2.0, 7.0] & 1.9   & 5.0   & 0.00\% &       & [0.1, 4.3] & 0.1   & 4.2   & 77.00\% &       & 17.70\% & 10.60\% \\
    XGBoost & 0.1   &       & [1.7, 7.1] & 1.6   & 5.4   & 0.00\% &       & [0.1, 4.5] & 0.1   & 4.4   & 74.90\% &       & 17.50\% & 10.10\% \\
    \midrule
    \multicolumn{15}{l}{$\sigma_{\varepsilon} = 1\quad n=2000$} \\
    Logit & 0.1   &       & [0.4, 17.6] & 0.3   & 17.3  & 10.90\% &       & [0.1, 18.2] & 0.0   & 18.2  & 79.40\% &       & 33.00\% & 20.40\% \\
    NBayes & 0.1   &       & [0.5, 4.5] & 0.4   & 3.9   & 0.20\% &       & [0.0, 2.9] & 0.0   & 2.9   & 96.00\% &       & 13.90\% & 10.40\% \\
    SVM   & 0.1   &       & [1.3, 7.1] & 1.2   & 5.8   & 0.00\% &       & [0.0, 4.0] & 0.0   & 4.0   & 97.20\% &       & 17.40\% & 11.90\% \\
    KNN   & 0.1   &       & [2.0, 16.1] & 1.9   & 14.1  & 0.00\% &       & [0.0, 10.1] & 0.0   & 10.1  & 95.20\% &       & 26.10\% & 16.90\% \\
    RF    & 0.1   &       & [0.6, 3.9] & 0.5   & 3.3   & 0.00\% &       & [0.0, 2.7] & 0.0   & 2.7   & 95.50\% &       & 13.50\% & 7.60\% \\
    XGBoost & 0.1   &       & [0.5, 3.7] & 0.4   & 3.2   & 0.00\% &       & [0.0, 2.3] & 0.0   & 2.3   & 96.00\% &       & 13.10\% & 7.00\% \\
    \midrule
    \multicolumn{15}{l}{$\sigma_{\varepsilon} = 2\quad n=500$} \\
    Logit & 1.9   &       & [1.6, 19.8] & 0.3   & 18.2  & 66.10\% &       & [0.3, 20.7] & 0.0   & 20.4  & 98.70\% &       & 38.00\% & 23.70\% \\
    NBayes & 1.9   &       & [2.3, 11.6] & 0.6   & 9.3   & 36.90\% &       & [0.2, 8.8] & 0.0   & 8.6   & 99.80\% &       & 24.10\% & 15.10\% \\
    SVM   & 1.9   &       & [3.0, 16.0] & 1.2   & 13.0  & 15.30\% &       & [0.3, 13.7] & 0.0   & 13.4  & 99.20\% &       & 31.10\% & 20.80\% \\
    KNN   & 1.9   &       & [1.8, 20.5] & 0.3   & 18.7  & 58.30\% &       & [0.3, 17.7] & 0.0   & 17.4  & 98.60\% &       & 35.10\% & 21.10\% \\
    RF    & 1.9   &       & [2.5, 10.8] & 0.7   & 8.3   & 27.50\% &       & [0.2, 8.0] & 0.0   & 7.8   & 99.80\% &       & 23.80\% & 14.10\% \\
    XGBoost & 1.9   &       & [2.3, 11.4] & 0.6   & 9.1   & 34.90\% &       & [0.2, 8.5] & 0.0   & 8.3   & 99.70\% &       & 24.50\% & 14.20\% \\
    \midrule
    \multicolumn{15}{l}{$\sigma_{\varepsilon} = 2\quad n=2000$} \\
    Logit & 1.9   &       & [0.3, 19.6] & 0.0   & 19.3  & 100.00\% &       & [0.1, 20.5] & 0.0   & 20.5  & 100.00\% &       & 35.70\% & 22.60\% \\
    NBayes & 1.9   &       & [0.5, 7.5] & 0.0   & 6.9   & 100.00\% &       & [0.0, 6.5] & 0.0   & 6.5   & 100.00\% &       & 19.80\% & 12.60\% \\
    SVM   & 1.9   &       & [1.5, 10.7] & 0.0   & 9.2   & 84.80\% &       & [0.0, 7.8] & 0.0   & 7.8   & 100.00\% &       & 23.10\% & 14.60\% \\
    KNN   & 1.9   &       & [1.7, 19.1] & 0.1   & 17.5  & 71.90\% &       & [0.0, 13.8] & 0.0   & 13.8  & 100.00\% &       & 30.60\% & 19.20\% \\
    RF    & 1.9   &       & [0.8, 7.4] & 0.0   & 6.7   & 100.00\% &       & [0.0, 6.3] & 0.0   & 6.3   & 100.00\% &       & 19.70\% & 11.00\% \\
    XGBoost & 1.9   &       & [0.9, 8.1] & 0.0   & 7.2   & 100.00\% &       & [0.0, 6.2] & 0.0   & 6.2   & 100.00\% &       & 20.60\% & 11.20\% \\
    \bottomrule
    \end{tabular}%
    }
  \label{tab:modelclibv-setup1}%
\end{table}%

\begin{table}[t!]
  \centering
  \caption{Monte Carlo simulation results on the interval estimation combined with model calibration under \textbf{Scenario 2}. All values except the percentages in the table are presented in units scaled by $10^{-2}$.}
  \resizebox{\textwidth}{!}{
    \begin{tabular}{lcrlccrrlccrrrr}
    \toprule
          &       &       & \multicolumn{4}{c}{Plug-in}   &       & \multicolumn{4}{c}{Partitioning} &       &       &  \\
\cmidrule{4-7}\cmidrule{9-12}    Method & $\theta$ &       & Estimate & Bias  & Width & CovR  &       & Estimate & Bias  & Width & CovR  &       & TMCR  & CMCR \\
    \midrule
    \multicolumn{15}{l}{$\sigma_{\varepsilon} = 1\quad n=500$} \\
    Logit & 19.7  &       & [3.6, 20.1] & 0.8   & 16.5  & 56.10\% &       & [1.9, 20.4] & 0.7   & 18.5  & 59.40\% &       & 39.30\% & 21.40\% \\
    NBayes & 19.7  &       & [14.1, 19.8] & 0.9   & 5.7   & 51.10\% &       & [17.2, 20.5] & 0.7   & 3.3   & 49.00\% &       & 23.10\% & 12.80\% \\
    SVM   & 19.7  &       & [8.7, 19.3] & 1.1   & 10.6  & 42.30\% &       & [10.6, 20.3] & 0.6   & 9.8   & 61.50\% &       & 32.40\% & 18.60\% \\
    KNN   & 19.7  &       & [3.8, 19.9] & 0.9   & 16.1  & 52.30\% &       & [7.3, 20.4] & 0.6   & 13.0  & 60.90\% &       & 35.90\% & 18.70\% \\
    RF    & 19.7  &       & [14.7, 19.4] & 1.0   & 4.7   & 43.50\% &       & [17.0, 20.5] & 0.7   & 3.5   & 51.20\% &       & 21.70\% & 10.60\% \\
    XGBoost & 19.7  &       & [15.7, 19.9] & 0.9   & 4.2   & 48.40\% &       & [17.6, 20.5] & 0.8   & 3.0   & 46.60\% &       & 20.00\% & 10.00\% \\
    \midrule
    \multicolumn{15}{l}{$\sigma_{\varepsilon} = 1\quad n=2000$} \\
    Logit & 19.7  &       & [3.1, 20.4] & 0.2   & 17.3  & 70.70\% &       & [2.0, 20.5] & 0.2   & 18.6  & 76.00\% &       & 37.10\% & 20.40\% \\
    NBayes & 19.7  &       & [17.7, 20.4] & 0.2   & 2.7   & 68.20\% &       & [18.7, 20.6] & 0.2   & 1.9   & 62.00\% &       & 19.10\% & 10.40\% \\
    SVM   & 19.7  &       & [13.7, 19.5] & 0.6   & 5.8   & 41.20\% &       & [16.1, 20.5] & 0.1   & 4.4   & 77.70\% &       & 24.50\% & 11.90\% \\
    KNN   & 19.7  &       & [6.2, 20.1] & 0.3   & 13.9  & 61.30\% &       & [12.0, 20.6] & 0.1   & 8.6   & 79.00\% &       & 30.60\% & 16.90\% \\
    RF    & 19.7  &       & [17.6, 20.2] & 0.2   & 2.7   & 67.70\% &       & [18.6, 20.6] & 0.2   & 2.0   & 65.00\% &       & 16.50\% & 7.60\% \\
    XGBoost & 19.7  &       & [18.2, 20.3] & 0.2   & 2.1   & 65.60\% &       & [18.8, 20.5] & 0.3   & 1.8   & 58.60\% &       & 16.30\% & 7.10\% \\
    \midrule
    \multicolumn{15}{l}{$\sigma_{\varepsilon} = 2\quad n=500$} \\
    Logit & 20.6  &       & [3.9, 22.0] & 0.5   & 18.1  & 69.30\% &       & [2.0, 22.2] & 0.4   & 20.3  & 71.80\% &       & 40.20\% & 23.60\% \\
    NBayes & 20.6  &       & [13.7, 21.7] & 0.6   & 8.1   & 65.60\% &       & [16.1, 22.4] & 0.3   & 6.3   & 74.50\% &       & 28.40\% & 15.20\% \\
    SVM   & 20.6  &       & [8.3, 21.2] & 0.7   & 13.0  & 58.30\% &       & [9.5, 22.3] & 0.3   & 12.8  & 75.20\% &       & 35.20\% & 20.60\% \\
    KNN   & 20.6  &       & [3.6, 21.9] & 0.5   & 18.3  & 66.20\% &       & [6.3, 22.3] & 0.4   & 16.0  & 74.10\% &       & 38.50\% & 21.00\% \\
    RF    & 20.6  &       & [13.3, 21.2] & 0.7   & 7.9   & 57.10\% &       & [15.8, 22.4] & 0.3   & 6.7   & 76.30\% &       & 28.60\% & 14.00\% \\
    XGBoost & 20.6  &       & [13.9, 21.8] & 0.5   & 7.9   & 66.50\% &       & [16.1, 22.4] & 0.3   & 6.3   & 73.00\% &       & 27.70\% & 14.30\% \\
    \midrule
    \multicolumn{15}{l}{$\sigma_{\varepsilon} = 2\quad n=2000$} \\
    Logit & 20.6  &       & [3.4, 22.7] & 0.0   & 19.3  & 92.60\% &       & [2.0, 22.8] & 0.0   & 20.8  & 94.70\% &       & 37.70\% & 22.60\% \\
    NBayes & 20.6  &       & [17.2, 22.6] & 0.0   & 5.4   & 94.00\% &       & [18.1, 22.8] & 0.0   & 4.8   & 95.10\% &       & 25.20\% & 12.60\% \\
    SVM   & 20.6  &       & [12.6, 21.7] & 0.1   & 9.1   & 78.20\% &       & [15.1, 22.8] & 0.0   & 7.7   & 96.40\% &       & 29.60\% & 14.60\% \\
    KNN   & 20.6  &       & [5.3, 22.4] & 0.1   & 17.1  & 91.10\% &       & [10.6, 22.9] & 0.0   & 12.2  & 95.70\% &       & 34.80\% & 19.20\% \\
    RF    & 20.6  &       & [16.8, 22.4] & 0.0   & 5.6   & 92.30\% &       & [18.0, 22.8] & 0.0   & 4.8   & 95.40\% &       & 24.20\% & 11.10\% \\
    XGBoost & 20.6  &       & [16.7, 22.4] & 0.0   & 5.7   & 93.00\% &       & [18.0, 22.8] & 0.0   & 4.8   & 94.90\% &       & 24.80\% & 11.30\% \\
    \bottomrule
    \end{tabular}%
    }
  \label{tab:modelclibv-setup2}%
\end{table}%

\newpage

\clearpage

\bibliographystyle{asa}
\bibliography{main}

\end{document}